\documentclass[12pt,preprint]{emulateapj}
\usepackage{natbib}
\usepackage{color}
\usepackage{hyperref} 
\hypersetup{colorlinks=true, citecolor=blue}

\def \bea {\begin{eqnarray}}
\def \ena {\end{eqnarray}}                  
\def \bee {\begin{equation}}
\def \ene {\end{equation}}
\def    \simlt  {\lower.5ex\hbox{$\; \buildrel < \over \sim \;$}}
\def    \simgt  {\lower.5ex\hbox{$\; \buildrel > \over \sim \;$}}

\newcommand     \mum    {\,\mu{\rm m}}  

\def	\cm		{\,{\rm {cm}}}
\def	\m		{\,{\rm m}}

\def	\B		{{\rm B}}
\def	\D		{{\rm D}}
\def	\erg		{\,{\rm {ergs}}}

\def    \exp 		{\,{\rm {exp}}}
\def	\g		{\,{\rm g}}

\def	\G		{\,{\rm G}}

\def	\K		{\,{\rm K}}
\def    \kB    		{k_{\rm B}}

\def	\s		{\,{\rm s}}

\def    \ln  		{\,{\rm {ln}}}
\def    \yr  		{\,{\rm {yr}}}

\def	\H		{\rm H}

\def	 \xhat		{\hat{\bf x}}
\def	 \yhat		{\hat{\bf y}}
\def	 \zhat		{\hat{\bf z}}
\def   \ahat 		{\hat{\bf a}}               
\def   \ehat 		{\hat{\bf e}}

\def	\V		{\rm V}

\def	\ehat		{\hat{\bf e}}

\def    \Bv     	{\bf  B}

\def	\ba		{{\bf a}}

\def	\bB		{{\bf B}}

\def	\bJ		{{\bf J}}

\def	\bM		{{\bf M}}

\def	\gas		{\rm {gas}}
\def	\th		{\rm {th}}

\def	\res		{\rm {res}}

\def	\d		{\rm d}
\def	\rad		{\rm {rad}}

\def    \ext    	{\rm {ext}}
\def    \pol    	{\rm {pol}}
\def    \DG		  {\rm {DG}}
\def    \Bar		  {\rm {Bar}}

\def	\eff		{\rm {eff}}
\def	\ali		{\rm {ali}}

\def	\gra	{\rm {gra}}
\def	\sil	  {\rm {sil}}
\def	\carb	{\rm {carb}}

\def	\mag 	{\rm {m}}

\def	\drag		{\rm {gas}}

\def	\Lar		{\rm {Lar}}
\def	\tot		{\rm {tot}}

\def	\LTE		{\rm {LTE}}

\def	\ISRF {\rm {ISRF}}

\def    \WMAP  		\it{ {WMAP}~}
\def    \Planck		\it{ {Planck}~}
\def    \ALMA		\it{ {ALMA}~}

\def	\obs		{\rm {obs}}
\def	\mod		{\rm {mod}}

\def	\ed		{\rm {ed}}
\def	\rot		{\rm {rot}}

\def	\C				{{\rm C}}

\font\mib=cmmib10

\def\bmu{\hbox{\mib\char"16}}

\def\bphi{\hbox{\mib\char"1E}}

\def\bomega{\hbox{\mib\char"21}}

\begin{document}
\shorttitle{Small grain alignment and UV polarization}
\shortauthors{Hoang, Lazarian, \& Martin}
\title{Paramagnetic alignment of small grains: A novel method for measuring interstellar magnetic fields}

\author{Thiem Hoang\altaffilmark{1,}\altaffilmark{2}, A. Lazarian\altaffilmark{3}, and P. G. Martin\altaffilmark{1}}
\affiliation{
$^1$ Canadian Institute for Theoretical Astrophysics, University of Toronto,
60 St. George Street, Toronto, ON M5S 3H8, Canada\\
$^2$Institut f$\ddot{\rm u}$r Theoretische Physik, Lehrstuhl IV: Weltraum- und 
Astrophysik, Ruhr-Universit$\ddot{\rm a}$t Bochum, 44780 Bochum, Germany\\
$^3$Department of Astronomy, University of Wisconsin, Madison, WI 53706, USA}

\begin{abstract}
We present a novel method to measure the strength of interstellar magnetic fields based on ultraviolet (UV) polarization of starlight, which is in part produced by weakly aligned, small interstellar grains. We begin with calculating degrees of alignment of small (size $a\sim 0.01\mum$) and very small ($a\sim 0.001\mum$) grains in the interstellar magnetic field due to the Davis-Greenstein paramagnetic relaxation and resonance paramagnetic relaxation. We compute the degrees of paramagnetic alignment with the ambient magnetic field $B$ using Langevin equations. In this paper, we take into account various processes essential for the dynamics of small grains, including infrared (IR) emission, electric dipole emission, plasma drag and collisions with neutral and ionized species. We find that the alignment of small grains is necessary to reproduce the observed polarization in the UV, although the polarization arising from these small grains is negligible at the optical and IR wavelengths. Based on fitting theoretical models to observed extinction and polarization curves, we find that the best-fit model requires a higher degree of alignment of small grains for the case with the peak wavelength of polarization $\lambda_{\max}<0.55\mum$, which exhibits an excess UV polarization relative to the Serkowski law, compared to the typical case $\lambda_{\max}=0.55\mum$. We interpret the correlation between the systematic increase of the UV polarization relative to maximum polarization (i.e. of $p(6\mum^{-1})/p_{\max}$) with $\lambda_{\max}^{-1}$ by appealing to the higher degree of alignment of small grains. We identify paramagnetic relaxation as the cause of the alignment of small grains and utilize the dependence of the degree of alignment on the magnetic field strength $B$ to suggest a new way to measure $B$ using the observable parameters $\lambda_{\max}$ and $p(6\mum^{-1})/p_{\max}$. Applying our new technique to the available observational data, we estimate the upper limit of interstellar magnetic field $B\sim 10-15\mu$G for the typical sightline ($\lambda_{\max}=0.55\mum$), which is consistent with the strength obtained from other available techniques. For the sightlines with lower $\lambda_{\max}$, the magnetic field strengths tend to be higher, assuming that the interstellar radiation field is similar along these sightlines. 
\end{abstract}

\section{Introduction}\label{sec61}

The polarization of starlight discovered more than a half century ago (\citealt{Hall:1949p5890}; \citealt{Hiltner:1949p5851}) revealed that interstellar dust grains must be nonspherical and aligned with respect to interstellar magnetic field. Since then, a number of alignment mechanisms have been proposed to explain why dust grains become aligned in the magnetic field, which include paramagnetic relaxation, mechanical torques, and radiative torques (see \citealt{2007JQSRT.106..225L} for a review).

In the present study we revisit the consequences of paramagnetic alignment, which was one of the first alignment mechanisms proposed to explain the polarization of starlight by \cite{1951ApJ...114..206D}. The mechanism relies on paramagnetic relaxation within rotating grains to align them with the magnetic field. Quantitative studies for the paramagnetic alignment mechanism (hereafter Davis-Greenstein (D-G) mechanism) were only conducted about two decades later. For instance, \cite{Jones:1967p2924} quantified the efficiency of the D-G mechanism using Fokker-Planck (FP) equations, while \cite{Purcell:1969p3641} and \cite{1971ApJ...167...31P} dealt with the problem by means of the Monte Carlo method. Their works showed that the D-G mechanism is inefficient for aligning grains in the typical interstellar magnetic field. Later, \cite{1979ApJ...231..404P} suggested that the joint action of pinwheel torques and paramagnetic relaxation would result in efficient alignment of suprathermally rotating grains (see also \citealt{Spitzer:1979p2708}).\footnote{The improved theory of paramagnetic alignment of suprathermally rotating grains is presented in \cite{1997ApJ...487..248L}.}

\cite{Lazarian:1997p5348} investigated analytically the D-G alignment for thermally rotating grains (grains not subject to pinwheel torques) while accounting for the Barnett relaxation effect (\citealt{Barnett:1915p6355}) and internal thermal fluctuations (\citealt{1994MNRAS.268..713L}; \citealt{1997ApJ...484..230L}). \cite{1999MNRAS.305..615R} (hereafter RL99) quantified the efficiency of the D-G mechanism for thermally rotating grains by numerically solving Langevin equations that describe the temporal evolution of grain angular momentum. These studies assumed a constant magnetic susceptibility $K(\omega)$ and considered the rotational damping and excitation of grains by gas atom bombardment. Also, the studies were performed for the $\sim 0.1\mum$ interstellar grains that rotate slowly ($\omega<10^{5}\s^{-1}$) not subject to spin-up systematic torques. The authors concluded that the D-G mechanism was inefficient to account for the dust polarization observed in molecular clouds, where the temperatures of dust and gas are expected to be comparable.

The first attempt to infer grain alignment from observations was performed by \cite{1995ApJ...444..293K}. The authors employed maximum entropy method to fit theoretical polarization curves to observational data and found that interstellar silicate grains of size $a \ge 0.05\mum$ (hereafter {\it typical} interstellar grains) are efficiently aligned while smaller grains are very weakly aligned. They found that there exists some residual alignment for $0.01\mum-0.05\mum$ grains (hereafter {\it small} grains, see also \citealt{2007EAS....23..165M}). Recently, in the interest of polarized submillimeter emission motivated by the {\it Planck} mission, \cite{Draine:2009p3780} derived the alignment function for interstellar grains by fitting simultaneously to the observed extinction and polarization curves for the typical diffuse ISM with $R_{V}=3.1$ and the peak wavelength (wavelength at maximum polarization) $\lambda_{\max}=0.55\mum$. They came to the same conclusion as \cite{1995ApJ...444..293K} that the typical interstellar grains are efficiently aligned. In addition, they found that the degree of alignment of small grains is $f \sim 0.01$ for the model with only silicate grains aligned. Due to its minor contribution to the polarization of starlight and to submillimeter/IR polarized emission, the problem of alignment of small grains was mostly forgotten. Through this study we argue that an in-depth understanding on the alignment of small grains can provide us new insight into the ISM.

\cite{1992ApJ...385L..53C} and \cite{1995ApJ...445..947C} reported an excess polarization in the UV from the Serkowski law (\citealt{Serkowski:1975p6681}) extrapolation for a number of stars with $\lambda_{\max}\le 0.53\mum$. {It is worth noting that the optical and IR polarization is mostly produced by typical interstellar grains aligned. The fact that the Serkowski law could fit well to the observational data from the optical to IR wavelength but failed for the UV \citep{1995ApJ...445..947C} reveals that the excess UV polarization should originate from some aligned grains that do not contribute to the optical and IR polarization. This could be a potential evidence for the alignment of small grains.}

\cite{1995ApJ...445..947C} found a tight correlation between the excess UV polarization, characterized by $p(6\mum^{-1})/p_{\max}$, and $\lambda_{\max}^{-1}$ for a number of stars. In particular, they showed no difference in the UV extinction for these stars, indicating that the properties of dust along these sightlines are not distinct from the general ISM. Using updated observational data, \cite{1999ApJ...510..905M} have confirmed the correlation between $p(6\mum^{-1})/p_{\max}$ and $\lambda_{\max}^{-1}$ and showed that the UV polarization can be described by a modified-Serkowski relation. A systematic change in the size distribution of aligned grains was suggested as a potential cause of the relationship (see \citealt{1999ApJ...510..905M}). \cite{2003JQSRT..79..881L} and \cite{2007JQSRT.106..225L} discussed the paramagnetic alignment of small grains and pointed out that it can provide upper limits on the interstellar magnetic field as strong magnetic fields in the diffuse ISM can overproduce the polarization from small grains distorting Serkowski relations. \cite{2007JQSRT.106..225L} argued that the paramagnetic alignment of small grains can be used for magnetic field studies. {\it In the absence of quantitative studies, however, the issue of the mechanisms responsible for the alignment of small grains remains open.}

Modern understanding on grain alignment has established a leading alignment mechanism based on radiative torques (RATs) induced by anisotropic radiation acting on realistic irregular grains. The mechanism was proposed by \cite{1976Ap&SS..43..291D} and numerically studied by \cite{1996ApJ...470..551D} and \cite{1997ApJ...480..633D}. The analytical model of RAT alignment was introduced in \cite{2007MNRAS.378..910L}. This model explained many puzzling features of the RAT alignment and provided the basis for quantitative predictions of the RAT alignment efficiencies. The theory was further elaborated in \cite{2008MNRAS.388..117H} and \cite{2009ApJ...697.1316H}, which improves its predictive abilities. Observational evidence for the RAT alignment was reported by a number of papers (\citealt{2007ApJ...665..369A}; \citealt{2008ApJ...674..304W}; \citealt{2011A&A...534A..19A}).

While the RAT alignment proves to be a robust alignment mechanism for large grains in various environment conditions, it appears to be inefficient for small grains due to the RAT magnitude decreasing with the decreasing grain size as $(\lambda/a)^{-\alpha}$ with $\alpha=3-4$ for $a\ll \lambda$ (\citealt{2007MNRAS.378..910L}). The spin-up mechanisms proposed by \cite{1979ApJ...231..404P} are also inefficient because the fast flipping of small grains tends to cancel out systematic torques fixed in the grain body (\citealt{1999ApJ...516L..37L}; \citealt{2009ApJ...695.1457H}). Therefore, a promising mechanism responsible for the weak alignment of small grains is the paramagnetic relaxation. As the degree of alignment by the paramagnetic mechanism depends strongly on the magnetic field strength, the grain alignment of small grains can open a new way to measure the ISM magnetic field based on the UV polarization. While the idea of using the UV polarization to measure magnetic fields was mentioned in \cite{2007JQSRT.106..225L}, no detailed study of the process has been carried out yet. {\it Quantitative study of paramagnetic alignment of small grains is the main goal of the present study.}

We should mention that special attention was paid to the alignment of very small ($a\sim 3-100$\AA) grains after the discovery of anomalous microwave emission (AME) (\citealt{Kogut:1996p5303}; \citealt{Leitch:1997p7359}). Electric dipole emission from very small, rapidly spinning grains (hereafter VSGs and polycyclic aromatic hydrocarbon (PAHs) are used interchangeably) is attributed to be the source of the AME (\citealt{1998ApJ...508..157D}). Recently, \cite{2010ApJ...715.1462H} and \cite{2011ApJ...741...87H} (hereafter HDL10 and HLD11)) have improved the original model of spinning dust emission by considering the emission from wobbling irregular grains subject to transient spin-up by single ion collisions, and the distribution of grain angular momentum is calculated exactly by solving Langevin equations. 

The polarization of spinning dust emission is of great interest to the Planck mission and other CMB B-mode polarization programs because the weak B-mode CMB signal requires very careful treatment of polarized Galactic foreground emissions. However, the question how ultrasmall grains are aligned and whether the spinning dust emission is polarized remains unclear. 

A problem with the classical treatment of paramagnetic relaxation, namely, its decrease of efficiency for rapidly rotating small grains, was addressed in \cite{2000ApJ...536L..15L} (hereafter LD00). LD00 found that the traditional treatment of paramagnetic relaxation is incomplete because it neglects the splitting of rotational energy levels in a rotating paramagnetic body. The authors pointed out that the Barnett effect that underlies the magnetization can allow the paramagnetic relaxation to occur resonantly at a maximum rate thanks to the splitting of rotational energy levels. This new effect that was termed resonance relaxation allowed LD00 to evaluate the alignment of VSGs. The degree of alignment reported in LD00 was less than $6\%$ but the predicted level of polarization was significant for high precision CMB polarization studies for which the polarization of AME acts as a foreground. In view of the advances in the description of the dynamics of wobbling irregular VSGs achieved in HDL10 and HLD11, it is timely to revisit the problem of alignment of the VSGs.

{ Observed extinction curves exhibit a prominent feature at $\lambda=2175$\AA. Such a UV bump is widely believed to originate from the electronic transition $\pi-\pi^{*}$ in $sp^{2}$-bonded carbon sheets of small graphite grains (\citealt{1965ApJ...142.1681S}; \citealt{1989IAUS..135..313D}) or PAHs (\citealt{2001ApJ...554..778L}; \citealt{2001ApJ...548..296W}). In the models by Draine and his co-workers, PAHs are suggested to be the dominant carrier of the $2175$\AA~ feature.}
\footnote{\cite{2003ssac.proc...37L} listed many other candidates that have been proposed as carriers of the $2175$\AA~ feature, including amorphous carbon, graphitized (dehydrogenated) hydrogenated amorphous carbon \citep{1986ApJ...305..817H}, nano-sized hydrogenated amorphous carbon \citep{1996ApJ...464L.187S}, quenched carbonaceous composite \citep{1995P&SS...43.1223S}, coals \citep{1995P&SS...43.1287P}, and OH$^{−}$ ion in low-coordination sites on or within silicate grains \citep{1989MNRAS.236..709D}.}  
Among about 30 stars for which the UV polarization data are observationally available to date, most of them do not show the polarization feature (bump) at $\lambda=2175$\AA~as seen in the extinction curves, except for two stars HD 197770 and HD 147933-4. This indicates that small carbonaceous grains may be only very weakly aligned. 

Taking advantage of the special UV polarization bumps at $2175$\AA~seen in HD 197770 and HD 147933-4, \cite{Hoang:2013dw} carried out the fitting to the observed data and inferred the alignment function for the entire range of grain size distribution. We found that the alignment of ultrasmall carbonaceous grains with the efficiency of $\sim 0.5\%$ is required to reproduce the $2175$\AA~ polarization bump of HD 197770. {\it The question now is that whether ultrasmall grains are aligned by the same mechanism as small grains.}

The goal of the present study is (1) to calculate the degree of alignment for small grains by the paramagnetic relaxation (e.g., D-G paramagnetic relaxation and resonance paramagnetic relaxation), taking into account various processes of rotational damping and excitation; (2) to derive the degree of alignment of small grains that reproduces observed polarization curves of the different $\lambda_{\max}$; and (3) to employ the inferred degree of alignment combined with the theoretical predictions to estimate the strength of interstellar magnetic fields. The paper is structured as follows.

In \S \ref{sec:timescales}, we describe the basic assumptions and principal dynamical timescales involved in the alignment problem. In \S \ref{sec:excitation} we briefly discuss major rotational damping and excitation processes and their diffusion coefficients. \S \ref{sec:D-G} is devoted to discussing the magnetic properties of dust grains and alignment mechanisms induced by the D-G paramagnetic relaxation and resonance paramagnetic relaxation. In \S \ref{sec:numeric}, we describe a numerical method to compute the degree of grain alignment using the Langevin equations and present the obtained results for both silicate and carbonaceous grains. In \S \ref{sec:obs} we present the alignment functions for small grains in the ISM inferred for the best-fit models to observed extinction and polarization. \S \ref{sec:Bfield} introduces a new technique to constrain the strength of magnetic field using UV polarization. Further discussion on the importance of our results and related effects and summary are presented in \S \ref{sec:discus} and \S \ref{sec:sum}, respectively.

\section{Assumptions and Dynamical Timescales}\label{sec:timescales}

\subsection{Grain geometry}
We consider oblate spheroidal grains with the moments of inertia $I_{1}>I_{2}=I_{3}$ along the grain's principal axes denoted by $\hat{\ba}_{1}$, $\hat{\ba}_{2}$ and $\hat{\ba}_{3}$. Let $I_{\|}=I_{1}$ and $I_{\perp}=I_{2}=I_{3}$. They take the following forms: 
\bea
I_{\|}=\frac{2}{5}Ma_{2}^{2}=\frac{8\pi}{15}\rho a_{1}a_{2}^{4},\label{eq:Ipar}\\\\
I_{\perp}=\frac{4\pi}{15}\rho a_{2}^{2}a_{1}\left(a_{1}^{2}+a_{2}^{2}\right),\label{eq:Iperp}
\ena
where $a_{1}$ and $a_{2}=a_{3}$ are the lengths of semimajor and semiminor axes of the oblate spheroid with axial ratio $r= a_{2}/a_{1}>1$, and $\rho$ is the grain material density. A frequently used parameter in the following, $h=I_{\|}/I_{\perp}$, is equal to
\bea
h=\frac{2a_{2}^{2}}{a_{1}^{2}+a_{2}^{2}}=\frac{2}{1+s^{2}},\label{eq:h}
\ena
where $s=1/r=a_{1}/a_{2}<1$.

The grain size $a$ is defined as the radius of a sphere of equivalent volume, which is given by
\bea
a=\left(\frac{3}{4\pi} (4\pi/3) a_{1}a_{2}^{2}\right)^{1/3}=a_{2}s^{1/3}.\label{eq:aeff}
\ena

\subsection{Barnett relaxation}\label{sec:Bar}
\cite{Barnett:1915p6353} first pointed out that a rotating paramagnetic body can get magnetized with the magnetic moment along the grain angular velocity.\footnote{This is an inverse of the Einstein-de Haas effect that was used to measure the spin of the electron.} Later, \cite{1976Ap&SS..43..257D} introduced the magnetization via the Barnett effect for dust grains and considered its consequence on grain alignment. 

The instantaneous magnetic moment due to the Barnett effect is equal to
\bea
\bmu_{\Bar}=\frac{\chi(0)\bomega}{\gamma_{g}}V=-\frac{\chi(0)\hbar V}{g_{e}\mu_{B}}\bomega,\label{eq:muBar}
\ena
where $V$ is the grain volume, $\gamma_{g}=-g_{e}\mu_{B}/\hbar\approx -e/(m_{e}c)$ is the gyromagnetic ratio of an electron, $g_{e}\approx 2$ is $g-$factor, $\mu_{\B}=e\hbar/2m_{e}c\approx 9.26\times 10^{-21} \erg \G^{-1}$ is the Bohr magneton. In the above equation, $\chi(0)$ is the zero-frequency paramagnetic susceptibility (i.e., at $\omega=0$), which reads
\bea
\chi(0)=4.2\times10^{-2}f_{p}\left(\frac{T_{\d}}{15\K}\right)^{-1},
\ena
where $T_{\d}$ is the grain temperature, and $f_{p}$ is the fraction of {\it paramagnetic} atoms (i.e., atoms with partially-filled shells) in the grain (see \citealt{Draine:1996p6977} and references therein). An extended discussion on the magnetic properties of interstellar dust is presented in \S \ref{sec:mag}.

\cite{1979ApJ...231..404P} realized that the precession of $\bomega$ coupled to $\bmu_{\Bar}$ around the grain symmetry axis $\ahat_{1}$ produces a rotating magnetization component within the grain body coordinates. As a result, the grain rotational energy is gradually dissipated until $\bomega$ becomes aligned with $\ahat_{1}$-- an effect which Purcell termed "Barnett relaxation". \cite{1999ApJ...520L..67L} (henceforth LD99a) revisited the problem by taking into account both spin-lattice and spin-spin relaxation (see \citealt{Morrish:1980}). Another internal relaxation process discussed in \cite{1979ApJ...231..404P} is related to imperfect elasticity of the grain material, which was expected to be important for grains of suprathermal rotation only (see e.g., \citealt{1997ApJ...484..230L}).

Following LD99a, \footnote{Due to a typo, the term $(\hbar/g_{n}\mu_{N})^{2}$ in Eq. 7 of LD99a should be replaced by $(\hbar/g_{e}\mu_{B})^{2}\equiv 1/\gamma_{g}^{2}$ which is applied for electron spins.} the Barnett relaxation time is defined as:
\bea
\tau_{\Bar}=\frac{\gamma_{g}^{2}I_{\|}^{3}}{VK(\omega_{1})h^{2}(h-1)J^{2}},\label{eq:tauBar}
\ena
where $K(\omega_{1})$ is related to the imaginary part of the magnetic susceptibility $\chi"$ as follows:
\bea
K(\omega_{1})&=&\frac{\chi''_{e}(\omega_{1})}{\omega_{1}}=\frac{\chi(0)\tau_{el}}{[1+(\omega_{1}\tau_{el}/2)^{2}]^{2}}\\
&\approx& \frac{1.2\times 10^{-13}\s}{[1+(\omega_{1}\tau_{el}/2)^{2}]^{2}}
\ena
where $\omega_{1}=(h-1)J\cos\theta/I_{\|}$ is the precession frequency of $\bomega$ around $\hat{\ba}_{1}$, and $\tau_{el}$ is the relaxation time of electronic spins.

For oblate spheroidal grains, we obtain
\bea
\tau_{\Bar}&\approx& 2.33\hat{\rho}^{2}a_{-5}^{7}\hat{s}^{-4/3}\left(\frac{1+s^{2}}{1.25}\right)^{2}\left(\frac{J_{\d}}{J}\right)^{2}\nonumber\\
&&\times \hat{K}^{-1}\left[1+\left(\omega_{1}\tau_{el}/2\right)^{2}\right]^{2}\yr,~~~\label{eq:316}
\ena
where $a_{-5}=a/10^{-5}\cm$, $\hat{s}=s/0.5$, $\hat{\rho}=\rho/3\g\cm^{-3}$, $\tau_{el}\sim \tau_{2}\sim 2.9\times 10^{-12}f_{p}^{-1}\s$ with assumption of $f_{p}=0.1$ is the spin-spin relaxation time, $\hat{K}=\chi(0)\tau_{el}/1.2\times 10^{-13}\s$, and $J_{\d}=\sqrt{I_{\|}k_{\B}T_{\d}/(h-1)}$ is the dust thermal angular momentum.\footnote{The relaxation of electronic spins results from the spin-lattice and spin-spin relaxation, with time scales $\tau_{1}\gg\tau_{2}$, so here we adopted $\tau_{el}\sim \tau_{2}$ (\citealt{Draine:1996p6977}).}

Although \cite{1979ApJ...231..404P} considered the grains having both electronic and nuclear spins, his study missed the effect of internal relaxation related to nuclear spins. LD99a found that for astrophysical grains of realistic composition nuclear spins induce a new type of relaxation, which was termed "nuclear relaxation" by LD99a. This relaxation process was shown to be dominant for large grains but it is negligible for small grains considered in this paper.

{ Internal relaxation involves the transfer of grain rotational energy to vibrational modes. Naturally, if the grain has nonzero vibrational energy, energy can also be transferred from the vibrational modes to grain rotational energy \citep{Jones:1967p2924}. For an isolated grain, a small amount of energy gained from the vibrational modes can induce fluctuations of the rotational energy $E_{\rot}$ when the grain angular momentum $\bJ$ is conserved (\citealt{1994MNRAS.268..713L}). Over time, the fluctuations in $E_{\rot}$ establish a local thermal equilibrium (LTE). 

Using the rotational energy of oblate spheroid $E_{\rot}=J^{2}\left[1+(h-1)\sin^{2}\theta\right]/2I_{\|}$, the fluctuations of the rotational energy can be described by the Boltzmann distribution \citep{1997ApJ...484..230L}:}
\bea
f_{\LTE}(J, \theta)= A{\exp}\left(-\frac{J^{2}}{2I_{\|}k_{\B}T_{\d}}
\left[1+(h-1)\sin^{2}\theta\right]\right),\label{eq:606}
\ena
where $A$ is a normalization constant such that $\int_{0}^{\pi} f_{\LTE}(J,\theta)\sin\theta d\theta=1$.

\subsection{Larmor precession of $\bJ$ around $\Bv$}

A rotating paramagnetic grain can acquire a magnetic moment due to the Barnett effect (Eq. \ref{eq:muBar}) and the Rowland effect if the grain is electrically charged \citep{1971MNRAS.153..279M}. The former is shown to be much stronger than that arising from the rotation of its charged body \citep{1976Ap&SS..43..291D}.

The interaction of the grain's magnetic moment due to the Barnett effect with an external static magnetic field $\bB$, governed by the torque $[\bmu_{\Bar}\times\bB]=-|\mu_{\Bar}|B\sin\beta\hat{\bphi}\equiv I_{\|}\omega\sin\beta d\phi/dt \hat{\bphi}$, causes the rapid precession of $\bJ\| \bomega$ around $\bB$. The period of such a Larmor precession denoted by $\tau_{\B}$, is given by
\bea
\tau_{\B}&=&\frac{2\pi}{d\phi/dt}=\frac{2\pi I_{\|}\omega}{|\mu_{\Bar}|B}=\frac{2\pi I_{\|}g\mu_{B}}{\chi_{0}V\hbar B}\nonumber\\
&\approx&1.32 a_{-5}^{2}\hat{s}^{-2/3}\hat{\rho}\hat{\chi}^{-1}
\hat{B}^{-1} ~\yr,
\label{eq:tauB}
\ena
where $\hat{B}=B/5 \mu\G$, $\hat{\chi}=\chi(0)/10^{-4}$.

\subsection{Measures of Alignment and Rayleigh Reduction Factor}

Let $G_{X}$ be the degree of alignment of the axis of major inertia $\hat{\ba}_{1}$ of the grain with its angular momentum $\bJ$ (i.e., internal alignment) and $G_{J}$ be the degree of alignment of $\bJ$ with the ambient magnetic field $\Bv$ (i.e., external alignment, see Figure \ref{fig:frames}). They are respectively given by
\bea
G_{X}&=&\frac{1}{2}\left(3\cos^{2}\theta-1\right),\label{eq:GX}\\
G_{J}&=& \frac{1}{2}\left(3\cos^{2}\beta-1\right).\label{eq:GJ}
\ena

Since we are interested in the mean alignment of an ensemble of grains with different orientations, the degrees of internal alignment and external alignment of grains are usually given by their ensemble averages, i.e., $Q_{X}=\langle G_{X}\rangle$ and $Q_{J}=\langle G_{J}\rangle$.

The net degree of alignment of the grain axis of major inertia with the magnetic field, namely Rayleigh reduction factor, is defined as
\bea
R=\langle G_{X}(\cos^{2}\theta)G_{J}(\cos^{2}\beta)\rangle.\label{eq:Rfactor}
\ena

In the regime of efficient Barnett relaxation, the fast variable $\theta$ can be separated from the slow variables $J$ and $\beta$ (\citealt{1997MNRAS.291..345R}). Therefore, the internal alignment can be described by the mean degree of alignment
\bea
q_{X}(J)=\int G_{X} f_{\LTE}(J,\theta)\sin\theta d\theta 
\ena
and the Raleigh reduction factor becomes
\bea
R=\int G_{J}(\cos^{2}\beta)q_{X}(J)  f(J_{x},J_{y},J_{z})d^{3}J , \label{eq:R}
\ena
where the distribution of grain angular momentum $f(\bJ)$ is used.

\section{Rotational Damping and Excitation Processes}\label{sec:excitation}
{ For typical and big interstellar grains, theoretical calculations show that the rotational damping by random collisions of the grain with gas atoms and molecules is dominant}. For small grains under interest, in addition to the gas collisions, the damping is caused by various processes, e.g., IR emission (\citealt{Purcell:1969p3641}), interactions with passing ions, electric dipole emission.

\cite{1998ApJ...508..157D} investigated in detail rotational damping and excitation processes for VSGs, including PAHs. They derived diffusion coefficients for planar PAHs rotating around its symmetry axis. HDL10 improved DL98 results and calculated the diffusion coefficients for planar PAHs with its rotation axis disaligned with grain angular momentum. Here we deal with the alignment of small grains and VSGs of oblate spheroidal shape.

\subsection{Rotational damping and excitation coefficients}
We follow the definitions of rotational damping $F$ and excitation coefficients $G$ from \cite{1998ApJ...508..157D}. The dimensionless damping coefficient for the $j$ process, $F_{j}$, is defined as the ratio of the damping rate induced by that process to that induced by the collisions of gas species, $\tau_{\H}^{-1}$, assuming that the gas consists of purely atomic hydrogen:
\bea
F_j=\left(-\frac{d\omega}{\omega dt}\right)_{j}\left(\frac{1}{\tau_{\H}^{-1}}\right)\label{eq:F}
\ena
and the excitation coefficient is defined as
\bea
G_j=\left(\frac{I d\omega^{2}}{2dt}\right)_{j}\left(\frac{\tau_{\H}}{\kB T_{\rm gas}}\right),\label{eq:G}
\ena
where $j$=n, i, p and IR denote the grain collisions with neutral and ion, plasma-grain interactions, and IR emission, $\left(I d\omega^{2}/2dt\right)_{j}$ is the rate of increase of kinetic energy for rotation along the axis that has moment of inertia $I$ due to the excitation process $j$, $T_{\rm gas}$ is the gas temperature. For an uncharged grain in a gas of purely atomic hydrogen, $F_{n}=1$ and $G_{n}=1$.

To calculate the damping and excitation coefficients for wobbling grains, we follow the same approach as in HDL10, where the parallel components $F_{j,\|}$ and $G_{j,\|}$, and perpendicular components $F_{j,\bot}$ and $G_{j,\bot}$ with respect to $\ahat_{1}$ are computed using the general definitions (Equations \ref{eq:F} and \ref{eq:G}). The only modification is the moments of inertia $I_{\|}$ and $I_{\perp}$, which are given by Equations (\ref{eq:Ipar}) and (\ref{eq:Iperp}) for oblate spheroid instead of those for disk-like grains in HDL10.

For example, the characteristic damping times of an oblate spheroidal grain with $s=a_{1}/a_{2}<1$ for rotation along the directions parallel and perpendicular to the grain symmetry axis $\ahat_{1}$ are respectively given by
\bea
\tau_{\H,\|}=\frac{3I_{\|}}{4\sqrt{\pi}n_{\H}m_{\H}v_{\th}a_{2}^{4}\Gamma_{\|}},\label{eq:tauHx}\\
\tau_{\H,\perp}=\frac{3I_{\perp}}{4\sqrt{\pi}n_{\H}m_{\H}v_{\th}a_{2}^{4}\Gamma_{\perp}},\label{eq:tauHy}
\ena
where $\tau_{\H,\|}\equiv\tau_{\H,z}, \tau_{\H,\perp}\equiv \tau_{\H,y}=\tau_{\H,x}$ with $z$ the grain symmetry axis, and $x$ and $y$ being the axes perpendicular to the symmetry axis (see \citealt{Lazarian:1997p5348}). In the above equation, $n_{\H}$ is the gas density, $m_{\H}$ is the hydrogen mass, $v_{\th}$ is the thermal velocity of hydrogen, and the geometrical factors $\Gamma_{\|}$ and $\Gamma_{\perp}$ were derived in \cite{1993ApJ...418..287R} and given in Appendix \ref{apdx:A}.

For the typical parameters of the ISM, Equations (\ref{eq:tauHx}) and (\ref{eq:tauHy}) become
\bea
\tau_{\H,\|}&\approx& 6.58\times 10^{4} \hat{\rho} \left(\frac{s}{0.5}\right)^{2/3} a_{-5}\nonumber\\
&& \times \left(\frac{n_{\H}}{30\cm^{-3}}\right)^{-1}\left(\frac{T_{\gas}}{100\K}\right)^{-1/2} \Gamma_{\|}^{-1}\yr~~~
\ena
and
\bea
\tau_{\H,\perp}&\approx &4.11\times 10^{4}\hat{\rho} \left(\frac{s}{0.5}\right)^{2/3}\left(\frac{1+s^{2}}{1.25}\right) a_{-5}\nonumber\\
&&\times \left(\frac{n_{\H}}{30\cm^{-3}}\right)^{-1}\left(\frac{T_{\gas}}{100\K}\right)^{-1/2} \Gamma_{\perp}^{-1}\yr
\ena

Likewise, the characteristic damping times due to the electric dipole emission from HDL10 can be rewritten as
\bea
\tau_{\ed,\|}=\frac{3I_{\|}c^{3}}{6k_{\B}T_{\gas}\mu_{\perp}^{2}},\label{eq:tauedz}\\
\tau_{\ed,\perp}=\frac{3I_{\perp}c^{3}}{6k_{\B}T_{\gas}\left(\mu_{\perp}^{2}/2+\mu_{\|}^{2}\right)},\label{eq:tauedxy}
\ena
where $\mu_{\|}$ and $\mu_{\perp}$ are the components of the electric dipole moment $\bmu$ parallel and perpendicular to the grain symmetry axis. Here we assume an isotropic distribution of $\bmu$, which corresponds to  $\mu_{\|}^{2}=\mu_{\perp}^{2}/2=\mu^{2}/3$ where $\mu^{2}$ is given by Equation (11) in \cite{1998ApJ...508..157D}.

\subsection{Relative importance of the different interaction processes}

Depending on environment conditions, the damping and excitation process by gas-dust interactions (i.e., collisions and plasma drag) or IR emission dominates. For small grains, in the hot diffuse ISM, including warm neutral medium (WNM), warm ionized medium (WIM), or in reflection nebula with strong radiation, the damping by IR emission is the most important process. In the cold neutral medium (CNM) and molecular clouds where gas density is higher and starlight photons are shielded, the damping by gas-dust interactions dominate. For ultrasmall grains (e.g., PAHs), electric dipole emission induces the most significant damping (see \citealt{1998ApJ...508..157D} for detailed discussion). 

Table \ref{ISM} presents physical parameters for idealized environments where $\chi=u_{\rm rad}/u_{\rm ISRF}$ is the ratio of radiation energy density
$u_{\rm rad}$ to the mean radiation density for the diffuse interstellar medium 
$u_{\rm ISRF}$ (see \citealt{1983A&A...128..212M}), 
$n({\rm H}_{2}),~ n({\rm H}^{+}),~ n({\rm M}^{+})$
are the molecular hydrogen density, ion hydrogen density and ionized metal
density, respectively.

\begin{table}[htb]
\begin{center}
\caption{Idealized Environments For 
 Interstellar Matter}\label{ISM}
\begin{tabular}{llll} \hline\hline\\
\multicolumn{1}{c}{\it Parameters} & \multicolumn{1}{c}{CNM}& 
{WNM} &{WIM}\\[1mm]
\hline\\
$n_{\H}$~(cm$^{-3}$) &30 &0.4 &0.1 \\[1mm]
$T_{\gas}$~(K)& 100 & 6000 &8000 \\[1mm]
$\chi$ &1 &1 &1 \\[1mm]
$x_{\H}=n(\H^{+})/n_{\H}$ &0.0012 &0.1 &0.99 \\[1mm]
$x_{\rm M}=n(\rm M^{+})/n_{\H}$ &0.0003 &0.0003 &0.001 \\[1mm]
$y=2n({\H}_{2})/n_{\H}$& 0. & 0. & 0. \cr
\cr
\hline\hline\\
\end{tabular}
\end{center}
\end{table}

\section{Paramagnetic Alignment mechanism for small grains}\label{sec:D-G}
\subsection{Davis-Greenstein Paramagnetic Relaxation}
A classical mechanism of grain alignment based on paramagnetic relaxation was proposed by \cite{1951ApJ...114..206D}. The underlying idea of the mechanism is that, a paramagnetic grain gets magnetized with an instantaneous magnetization $\bM$ parallel to the induced magnetic field.
If the grain angular momentum makes an angle $\beta$ with $\Bv$, then $\Bv$ can be decomposed into the parallel $\Bv_{\|}$ and perpendicular $\Bv_{\perp}$ to $\bJ$. Since the paramagnetic material gets magnetized instantaneously in response to the induced magnetic field, the magnetization component $\bM_{\|}$ parallel to $\bJ$ remains constant during the grain rotation, while the perpendicular component $\bM_{\perp}$, fixed to the lab system, is rotating with respect to the grain body. As a result, the rotating magnetization experiences energy dissipation, which results in the gradual alignment of $\bJ$ with $\Bv$.

Due to the magnetic dissipation, the angle between $\bJ$ and $\Bv$ decreases as
\bea
I_{\|}\omega\frac{d\beta}{dt}=-K(\omega)VB^{2}\omega\sin\beta\cos\beta,\label{eq:dbeta_dt}
\ena
where $K(\omega)=\chi''(\omega)/\omega$ with $\chi"(\omega)$ being the imaginary part of complex magnetic susceptibility of the grain material at the rotation frequency $\omega$. In deriving the above equation, $\bomega$ and $\hat{\ba}_{1}$ are assumed to be aligned with $\bJ$ due to fast internal relaxation.

Equation (\ref{eq:dbeta_dt} can be rewritten as
\bea
\frac{d\beta}{dt}=-\frac{\sin\beta\cos\beta}{\tau_{\rm DG}},\label{eq:dbeta_tauD-G}
\ena
where
\bea
\tau_{\rm DG}=\frac{I_{\|}}{K(\omega)VB^{2}}
\ena 
is the characteristic timescale of paramagnetic alignment.

For the normal paramagnetic material, $\tau_{\DG}$ can be written as
\bea
\tau_{\rm DG}&=&\frac{2\rho a_{2}^{2}}{5K(\omega)B^{2}}\nonumber\\
&\approx& 2.0\times 10^{6}\hat{\rho}\hat{s}^{-2/3}a_{-5}^{2}\left(\frac{B}{5\mu\rm G}\right)^{-2}
\left(\frac{1.2\times10^{-13}\s}{K(\omega)}\right)\yr.~~~~~~\label{eq:tau_D-G}
\ena

\cite{Jones:1967p2924} employed the Fokker-Planck equations to compute the degree of alignment of angular momentum $Q_{J}$ in the magnetic field subject to the gas atom bombardment. Their obtained value $Q_{J}$ is equal to
\bea
Q_{J}&=\frac{3}{2}q(x),\label{eq:QJtheor}
\ena
where
 \bea
 x=\left(\frac{T_{\rm av}}{T_{\gas}}-1\right)=\left(\frac{\delta}{1+\delta}\times \frac{T_{\d}-T_{\gas}}{T_{\gas}}\right)
 \ena
with $\delta=\tau_{\drag}/\tau_{\DG}$ and $\tau_{\gas}=\tau_{\H,\|}$. Here $T_{\rm av}$ is regarded as the rotational temperature, and $q(x)$ takes the following form:
\bea
q(x)=-\frac{1}{3}+\frac{1}{x}\left[\left(\frac{1+x}{x}\right)^{1/2}{\rm arcsinh}\sqrt{x}-1\right],
\ena
for $x>0$. For $x<0$, the term $x^{-1/2}{\rm arcsinh}\sqrt{x}$ is replaced by
$(-x)^{-1/2}{\rm arcsin}\sqrt{-x}$, hence
\bea
q(x)=-\frac{1}{3}+\frac{1}{x}\left[\left(\frac{1+x}{-x}\right)^{1/2}{\rm arcsin}\sqrt{-x}-1\right].
\ena

The degree of the internal alignment for the case $T_{\d}=T_{\gas}$ (i.e., the distribution of angular momentum is Maxwellian) is equal to (\citealt{Jones:1967p2924}; \citealt{1997ApJ...484..230L})
\bea
Q_{X,\rm Mw}=\frac{3}{2\left(1-h^{-1}\right)}\left[1-\frac{1}{\sqrt{h-1}}{\rm arcsin}(1-h^{-1})\right]-\frac{1}{2}.~~~~~
\ena

Using Monte Carlo simulations, \cite{1971ApJ...167...31P} showed that Equation (\ref{eq:QJtheor}) provides a good agreement with their numerical calculations. The paramagnetic alignment of oblate grains was studied analytically in \cite{Lazarian:1997p5348} and numerically in RL99, accounting for the Barnett relaxation.

\subsection{Magnetic properties of interstellar dust}\label{sec:mag}

Following \cite{1999ApJ...512..740D}, the critically damped susceptibility  is given by $\chi"(\omega)=\omega K(\omega)$ with
\bea
K(\omega)=\frac{\chi(0)\tau_{2}}{[1+\left(\omega\tau_{2}/2\right)^{2}]^{2}},\label{eq:Kappa}
\ena
where $\chi(0)$ is the magnetic susceptibility at the zero rotation frequency. Using the Curie's law for paramagnetic material, we have
\bea
\chi(0)&=&\frac{n_{p}\mu^{2}}{3k_{\B}T_{\d}},\label{eq:curielaw}
\ena
where the effective magnetic moment $\mu$ reads
\bea
\mu^{2}\equiv p^{2}\mu_{\B}^{2} =g_{e}^{2}\mu_{B}^{2}\left[{J(J+1)}\right]
=\gamma_{g}^{2}\left[\hbar^{2}J(J+1)\right],\label{eq:mu}
\ena
with $J$ being the angular momentum quantum number of electrons in the outer partially filled shell and $p\approx 5.5$ (see \citealt{Draine:1996p6977}).\footnote{\cite{Draine:1996p6977} presented the total magnetic moment as $\mu=p\mu_{B}$ with $p=5.9$. One can see that for Fe$^{3+} (6S_{5/2})$ ion with $S=5/2, L=0$  and $J=5/2$ and $g_{e}\approx 2$, one obtain $g_{e} J(J+1)\equiv p=5.9$.} 
  
In Equation (\ref{eq:Kappa}), $\tau_{2}$ is the spin-spin relaxation time, which is equal to the precession time of the grain magnetic moment $\bmu$ around the magnetic field $H_{i}=3.8n_{p}\mu_{B}$:
\bea
\tau_{2}&=&\frac{\hbar}{g_{e}p\mu_{B}H_{i}}\approx \frac{\hbar}{3.8n_{p}g_{e}p\mu_{B}^{2}}\nonumber\\
&\approx& 2.9\times 10^{-11}\left(\frac{0.1}{f_{p}}\right)\left(\frac{10^{23}\cm^{-3}}{n_{\tot}}\right) \s,\label{eq:tau2}
\ena
where $n_{p}=f_{p}n_{\tot}$ is the number density of paramagnetic atoms and $n_{\tot}\approx 10^{23}\cm^{-3}$ is the total atomic number density within the grain (\citealt{Draine:1996p6977}). 
 
{\it Amorphous silicate grains} usually contain Si, Mg, Fe, and O atoms. Assuming the silicate material with structure MgFeSiO$_{4}$ containing Fe$^{3+}$ ($^{6}S_{5/2}$), the fraction of paramagnetic atoms is $f_{p}=1/7\approx 0.1$. The magnetization is induced by electrons in the outer partially filled shell of Fe$^{3+}$ ion having the structure $^{6}$S$_{5/2}$. Using $\gamma_{g}(\rm Fe)=-g_{e}\mu_{B}/\hbar=-1.76\times10^{7}\s^{-1}\G^{-1}$, one can estimate the static magnetic susceptibility for silicate grains as follows: 
\bea
\chi_{\sil}(0) \approx 4.2\times 10^{-3}\left(\frac{f_{p}}{0.1}\right)\left(\frac{p}{5.5}\right)^{2}
\left(\frac{n_{\tot}}{10^{23}\cm^{-3}}\right)\left(\frac{T_{\d}}{15\K}\right)^{-1}.\label{eq:chi0}
 \ena
Plugging in Equation (\ref{eq:chi0}) into (\ref{eq:Kappa}), one obtain
\bea
K_{\sil}(\omega)\approx 1.2\times 10^{-13}\left(\frac{T_{\d}}{15\K}\right)^{-1}\frac{1}{[1+\left(\omega\tau_{2}/2\right)^{2}]^{2}}
\s.\label{eq:Kappa_sil}
\ena

From Equation (\ref{eq:tau2}) and (\ref{eq:Kappa_sil}) one can see that for typical interstellar grains ($a>0.05\mum$) rotating with $\omega\sim \omega_{\th}=\left(2k_{\B}T_{\gas}/I_{\|}\right)^{1/2}=1.85\times 10^{5}a_{-5}^{-5/2}\hat{T}_{\gas} \s^{-1}$, the term $\omega\tau_{2}\ll 1$. Thus, it is disregarded in earlier studies on paramagnetic alignment of interstellar grains (e.g., \citealt{Lazarian:1997p5348}; \citealt{1999MNRAS.305..615R}). On the other hand, small grains ($a\le 0.05\mum$) are expected to spin rapidly with $\omega>10^{5}\s^{-1}$. Thus, the term $\omega\tau_{2}$ becomes important, and the paramagnetic relaxation is suppressed due to the decrease of $K(\omega)$. For VSGs that rotate extremely fast of $\omega >10^{9}\s^{-1}$, $K(\omega)$ is substantially reduced. Thus, VSGs cannot be aligned by the classical D-G paramagnetic relaxation.

For {\it ultrasmall carbonaceous grains or PAHs}, the magnetization arises from the presence of free radicals, paramagnetic carbon rings, and captured ions (see \citealt{2000ApJ...536L..15L} and references therein). Following \cite{2000ApJ...536L..15L}, we take $f_{p}=0.01$ corresponding to $n_{p}=10^{21}\cm^{-3}$ for the typical atom number density $n_{\tot}=10^{23}\cm^{-3}$.

For {\it graphite grains}, known as diamagnetic material, the magnetization originates from the attachment of H atoms to the grain through hydrogenation. Since an H electron is already used to make a covalent bond with a C atom, the magnetization is only produced by the H nucleus (proton). The gyromagnetic ratio for the H nucleon is $\gamma_{g}(\H)=g_{n}\mu_{N}/\hbar\approx 2.67\times 10^{4}\s^{-1}\G^{-1}$ where $g_{n}\approx 5.59$ and $\mu_{N}=e\hbar/2m_{p}c\approx 5.04\times 10^{-24}\erg \G^{-1}$, which is three orders of magnitude smaller than that of a Fe atom present in silicate grains. Plugging in $J=1/2$ and $\gamma_{g}(\H)$ into Equation (\ref{eq:curielaw}) we obtain
\bea
\chi_{\gra}(0)\approx 9.6 \times 10^{-10}\left(\frac{f_{p}}{0.1}\right)
\left(\frac{n_{\tot}}{10^{23}\cm^{-3}}\right)\left(\frac{T_{\d}}{15\K}\right)^{-1}~~~~,\label{eq:chi0_carb}
\ena
where $f_{p}$ is the fraction of H atoms. If $f_{p}(\H)$ is too small ($\ll 0.1$), the magnetization becomes dominated by the nuclei of $^{13}\C$ that has $f_{p}(^{13}\C)\approx 0.01$ (see also LD99a).

The function $K(\omega)$ for graphite grains is given by Equation (\ref{eq:Kappa}) but the spin-spin relaxation time $\tau_{2}$ now is replaced by the nuclear relaxation time $\tau_{n}$ with $\tau_{n}^{-1}=\tau_{ne}^{-1}+\tau_{nn}^{-1}$. Following LD99a, $\tau_{ne}$ and $\tau_{nn}$ are given by
\bea
\tau_{ne}&=&\frac{\hbar g_{e}}{3.8n_{e}g_{n}^{2}\mu_{\rm N}^{2}}\approx 3\times 10^{-4}
\left(\frac{2.7}{g_{n}}\right)^{2}\left(\frac{10^{22}\cm^{-3}}{n_{e}}\right)\s,~~~~\\
\tau_{nn}&=& \frac{\hbar}{3.8g_{n}n_{n}\mu_{\rm n}^{2}}\approx0.58\tau_{ne}\left(\frac{n_{e}}{n_{n}}\right).
\ena
Plugging in the above equation into Equation (\ref{eq:Kappa}), one obtain
\bea
K_{\gra}(\omega)\approx 1.1\times 10^{-13}\left(\frac{T_{\d}}{15\K}\right)^{-1}\frac{1}
{[1+\left(\omega\tau_{n}/2\right)^{2}]^{2}}\s,~~~\label{eq:Kappa-gra}
\ena
for $n_{e}=n_{n}=f_{p}n_{\tot}=10^{22}\cm^{-3}$, assuming $f_{p}=f_{p}(\H)=0.1$.\footnote{\cite{Jones:1967p2924} suggested that due to nuclear paramagnetism, the lower bound for interstellar grains $K(\omega)\sim 10^{-12}/T_{\d}\s$ regardless of their composition.}

Since $\tau_{n}\gg \tau_{2}$, one can see that $K_{\gra}(\omega)\ll K_{\sil}(\omega)$. Indeed, for a grain of $a=10^{-5}\cm$ rotating at the thermal velocity $\omega_{\th} \sim 10^{5}\s^{-1}$, Equation (\ref{eq:Kappa-gra}) yields $K_{\gra}(\omega_{\th})\approx 10^{-18}\s$, compared to $K_{\sil}(\omega_{\th})\approx 10^{-13}\s$ for silicate grains. Thus, the paramagnetic alignment of graphite grains is rather inefficient.

\subsection{Resonance Paramagnetic Relaxation}
The traditional treatment of the paramagnetic magnetization by the Barnett effect within a rotating body relies on the following assumption: the magnetization within a rotating body in a static magnetic field is equivalent to the magnetization of a body at rest in a rotating ambient magnetic field. This assumption was adopted in \cite{1951ApJ...114..206D}. LD00 realized that the above treatment of paramagnetic relaxation is not exact because it neglects the splitting of rotational energy levels. They pointed out that the Barnett effect can help the paramagnetic dissipation to occur resonantly at a maximum rate thanks to the splitting of energy levels. Such a new effect, termed by LD00 the resonance relaxation, can occur whenever the grain rotates in the ambient magnetic field.

Assuming the critically damped balance (\citealt{1999ApJ...512..740D}), LD00 found that
\bea
K(\omega)=\frac{\chi(0)\tau_{2}}{1+\gamma^{2}g_{e}^{2}\tau_{1}\tau_{2}
H_{1}^{2}\sin^{2}\theta},\label{eq:Kappares}
\ena
where $\gamma=e/2m_{e}c$ (e.g., $\gamma=\gamma_{e}/g_{e}$), $\tau_{1}$ is the spin-lattice relaxation time. Their estimate yields
\bea
\gamma^{2}g_{e}^{2}\tau_{1}\tau_{2}H_{1}^{2}\sin^{2}\theta
\approx 8\left(\frac{\tau_{1}}{10^{6}\s}\right)\left(\frac{\tau_{2}}{2\times 10^{-9}\s}\right)\left(\frac{H_{1}}{5\mu\G}\right)^{2}\left(\frac{\sin^{2}\theta}{2/3}\right).
\ena

Following LD00, the spin-lattice relaxation time of dust grains at a temperature $T_{\d}$, $\tau_{1}(T_{\d})$ is given by
\bea
\frac{\tau_{1}(T_{\d})}{\tau_{1,\infty}(77\K)}\approx \left(\frac{77\K}{T_{\d}}\right)^{m+1}\left(\frac{T_{\d}}{T_{l}}\right)^{m}
\exp\left(\frac{T_{\d}}{T_{l}}\right)m!\zeta(m),~~~\label{eq:tau1}
\ena
where $\zeta(m)$ is the Riemann zeta function for $m=6$ or $m=8$, and $T_{l}$ is the lowest grain vibrational temperature, which is equal to
\bea
T_{l}=\frac{\hbar\omega_{\min}}{k_{\B}}\approx 63\left(\frac{10^{-7}\cm}{a}\right) \K,
\ena
and $\tau_{1,\infty}(77\K)\approx 10^{-6}$ for the spin-lattice relaxation.

The uncertainty of the resonance relaxation arises from uncertainties of the microphysics of spin-lattice relaxation within VSGs. For such grains, LD00 used plausible arguments, but the laboratory testing would be most useful.

The timescale of magnetic alignment due to the D-G and resonance paramagnetic relaxation is equal to
\bea
\tau_{\mag}=\min \left(\tau_{\DG},\tau_{\res}\right),\label{eq:tau_mag}
\ena
where $\tau_{\DG}$ and $\tau_{\res}$ are obtained by plugging in $K(\omega)$ from Equations (\ref{eq:Kappa}) and (\ref{eq:Kappares}) into Equation (\ref{eq:tau_D-G}), respectively.

\section{Numerical calculations of degree of paramagnetic alignment}\label{sec:numeric}
\subsection{Numerical Method}

RL99 have statistically calculated the efficiency of D-G alignment mechanism for dust grains using the Langevin equations, which was first suggested for studies of grain dynamics in \cite{1993ApJ...418..287R}. RL99 also took into account the Barnett effect and internal thermal fluctuations. Here, we study the paramagnetic alignment using the same approach as in RL99 but account for a variety of damping and excitation processes that are important for small grains, including the dust-gas collisions, IR emission, plasma drag, and electric dipole emission (see HDL10, HLD11).

Following HDL10, to study the alignment of the grain angular momentum $\bJ$ with the ambient magnetic field $\Bv$, we solve Langevin equations for the evolution of $\bJ$ in time in an inertial coordinate system using the Euler-Maruyama algorithm. The inertial coordinate system is denoted by $\ehat_{1}\ehat_{2}\ehat_{3}$ where $\ehat_{1}$-axis is chosen to be parallel to $\Bv$. The Langevin equations (LEs) read 
\bea
dJ_{i}=A_{i}dt+\sqrt{B_{ii}}dq_{i}\mbox{~for~} i=~x,~y,~z,\label{eq:612}
\ena
where $dq_{i}$ are the random variables generated from a normal distribution with zero mean and variance $\langle dq_{i}^{2}\rangle=dt$, $A_{i}=\langle {\Delta J_{i}}/{\Delta t}\rangle$ and $B_{ii}=\langle \left({\Delta J_{i}}\right)^{2}/{\Delta t}\rangle$ are drifting (damping) and diffusion coefficients defined in the inertial coordinate system.

The drifting and diffusion coefficients in the reference system fixed to the grain body, $A_{i}^{b}$ and $B_{ij}^{b}$, are related to the damping and excitation coefficients as follows:
\bea
A_{i}^{b}&=&-\frac{J_{i}^{b}}{\tau_{\gas,i}}=-\frac{J_{i}^{b}}{\tau_{\H,i}}F_{\tot,i},\\
B_{zz}^{b}&=&B_{\|}=\frac{2I_{\|}\kB T_{\rm gas}}{\tau_{\rm H,\|}}G_{\rm tot,\|},\\
B_{xx}^{b}&=&B_{yy}^{b}=B_{\perp}=\frac{2I_{\perp}\kB T_{\rm gas}}{\tau_{\rm H,\perp}}G_{\rm tot,\perp},
\ena
where  $F_{\tot,i}$ and $G_{\tot, ii}$ for $i=x,y,z$ (or $\perp, \|$) are the total damping and excitation coefficients from various processes which are defined by Equations (\ref{eq:F}) and (\ref{eq:G}), and $\tau_{\gas, i}=F_{\tot,i}/\tau_{\H,i}$. Using the transformation of diffusion coefficients from the body system $\ahat_{1}\ahat_{2}\ahat_{3}$ to the inertial system $\ehat_{1}\ehat_{2}\ehat_{3}$ (see Appendix \ref{sec:Bdiff}), we obtain the drifting and diffusion coefficients $A_{i}$ and $B_{ii}$ in the inertial system.

To account for the magnetic alignment, we need to add a damping term $-{J_{x,y}}/{\tau_{\mag}}$ to the drifting coefficient $A_{x,y}$ and an excitation term $B_{{\mag},xx}=B_{{\mag},yy}$ to the diffusion coefficient $B_{xx}$ and $B_{yy}$ (see Appendix \ref{sec:Bm}).

In dimensionless units, $J'\equiv J/I_{\|}\omega_{T,\|}$ with $\omega_{T,\|}\equiv \left(2k_{\B}T_{\gas}/I_{\|}\right)^{1/2}$ being the thermal angular velocity of the grain along the grain symmetry axis, and $t'\equiv t/\tau_{\H,\|}$, Equation (\ref{eq:612}) becomes 
\bea
dJ'_{i}=A'_{i}dt'+\sqrt{B'_{ii}}dq'_{i} \mbox{~for~} i= x,~y,~z,\label{eq:613}
\ena
where $\langle dq_{i}^{'2}\rangle=dt'$ and
\bea
A'_{i}&=&-\frac{J'_{i}}{\tau'_{\gas,{\eff}}}-\frac{2}{3}\frac{J_{i}^{'3}}
{\tau'_{\ed,{\eff}}}-\frac{J'_{i}}{\tau'_{\mag}}(1-\delta_{zi}),\label{eq:Ai}~~~~\\
B'_{ii}&=&\frac{B_{ii}}{2I_{\|}\kB T_{\gas}}\tau_{\H,\|}+\frac{T_{\d}}{T_{\gas}}\delta_{\mag}(1-\delta_{zi}),\label{eq:Bii}
\ena
where $\delta_{\mag}=\tau_{\H,\|}/\tau_{\mag}$, $\delta_{zi}=1$ for $i=z$ and $\delta_{zi}=0$ for $i\ne z$, 
\bea
\tau'_{\gas,{\eff}}= \frac{\tau_{\gas,{\eff}}}{\tau_{\H,\|}},~\tau'_{\ed,{\eff}}&=&\frac{\tau_{\ed,{\eff}}}{\tau_{\H,\|}},~~~
\ena
where $\tau_{\gas,{\eff}}$ and $\tau_{\ed,{\eff}}$ are the effective damping times due to dust-gas interactions and electric dipole emission that result from transforming damping coefficients $A_{i}$ from the body system to the inertial system (see Eq. E4 in HDL10). 

Equations (\ref{eq:613}) together with (\ref{eq:Ai}) and (\ref{eq:Bii}) are solved by the iterative method for $N_{\rm step}$ with the time step $dt'$. As in HDL10, we choose $dt'= 0.1\min[1/F_{\tot,\|}, 1/G_{\tot,\|}, \tau_{\rm ed,\|}/\tau_{\H,\|}]$ and $N_{\rm step}=10^{6}$ for all calculations. If the returning timestep $dt'>10^{-3}$, then we take $dt'=10^{-3}$.\footnote{For ultrasmall grains of $4$\AA, the damping time by electric dipole emission dominates with $\tau_{\H}/\tau_{ed}\sim 10^{2}$ (see e.g., HDL10). At this size, the resonance paramagnetic alignment occurs over $\tau_{\D-G,\rm res}\sim \tau_{\H}$. Therefore, the chosen timestep $dt'$ remains valid for solving the Langevin equations.} At each time step, the angular momentum $\bJ$ and the angle $\beta$ between $\bJ$ and $\Bv$ obtained from the Langevin equations are employed to compute the degrees of grain alignment.

Indeed, at each time step, the components of angular momentum $J_{x}, J_{y}$ and $J_{z}$ are computed. Then, we calculate $J$ and the angle $\beta$ between $\bJ$ and $\Bv$ such as $\cos\beta=J_{z}/J$. We can calculate $R$ as follows: 
\bea
 R\equiv \sum_{l=0}^{N_{\rm step}-1} \frac{G_{X}(\cos^{2}\theta)G_{J}(\cos^{2}\beta)}{N_{\rm step}}.
 \ena
 
Above, the angle $\theta$ is kept unchanged during the time interval of $dt'$, which is invalid when the fast internal relaxation is assumed. Therefore, $G_{X}$ would be replaced by $q_{X} (J)$. 

In practical, the actual value $R$ and its approximation $Q_{X}Q_{J}$ have some correlation, which can be described by
 \bea
R=\langle G_{X} G_{J} \rangle =Q_{X}Q_{J}\left(1+f_{\rm corr}\right),
 \ena
where $f_{\rm corr}$ is a correlation factor (see RL99). The case $f_{\rm corr}=0$ corresponds to no correlation, i.e., $\theta$ and $\beta$ are completely independent.
 
\subsection{Davis-Greenstein Alignment of Thermally Rotating Grains}
We first study the paramagnetic alignment of grains subject to a single rotational damping and excitation process by gas bombardment as in RL99. In this case, grains are expected to be rotating at thermal velocity.

\subsubsection{Alignment with constant $K(\omega)$}
As in RL99, we assume the magnetic susceptibility $K(\omega)$ to be constant by disregarding the term containing $\omega$ in Equation (\ref{eq:Kappa_sil}). This assumption is valid for typical interstellar grains that rotate thermally at $\omega \ll 2\tau_{2}^{-1}$. We consider two cases of low ($T_{\d}=4\K$) and normal ($T_{\d}=20\K$) grain temperature and a variety of the magnetic field strength $B$ for the CNM (see Table \ref{ISM} for more physical parameters). Oblate spheroidal grains with axial ratio $r=2$ and $r=1.5$ are adopted, and $f_{p}=0.1$ is taken for silicate material.

\begin{figure}
\centering
\includegraphics[width=0.4\textwidth]{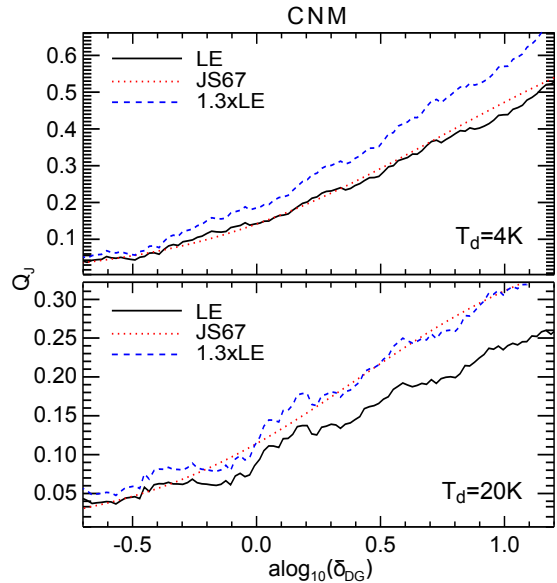}
\caption{Degree of grain alignment by the paramagnetic (D-G) relaxation for thermally rotating grains and constant $K(\omega)$ in the CNM as a function of $\delta_{\mag}=\tau_{\gas}/\tau_{\DG}$ for $T_{\d}=4\K$ (upper panel) and $T_{\d}=20\K$ (lower panel). The solid lines show our numerical results obtained by solving LEs, and dotted lines show the analytical results predicted for $T_{\d} \rightarrow 0 K$ from \cite{Jones:1967p2924}.}
\label{fig:QJ_RL99}
\end{figure}

Figure \ref{fig:QJ_RL99} shows our obtained results for $Q_{J}$ as a function of $\delta_{\mag}=\tau_{\gas}/\tau_{\DG}$. $Q_{J}$ appears to increase with the increasing $\delta_{\mag}$ as expected. The analytical results from \cite{Jones:1967p2924} for spherical grains are similar to our numerical results in the case $T_{\d}=4\K$. As $T_{\d}$ increases to $T_{\d}=20\K$, our numerical result is a factor of $1.3$ lower than the analytical prediction. This originates from the fact that $\chi(0)$ decreases when the thermal fluctuations within the grain (i.e., $T_{\d}$) increase.

\subsubsection{Effect of fast rotation}

When the grain rotation frequency becomes comparable to $2\tau_{2}^{-1}$, $K(\omega)$ decreases sharply according to Equation (\ref{eq:Kappa_sil}), resulting in the decrease of the paramagnetic alignment rate.

\begin{figure}
\centering
\includegraphics[width=0.4\textwidth]{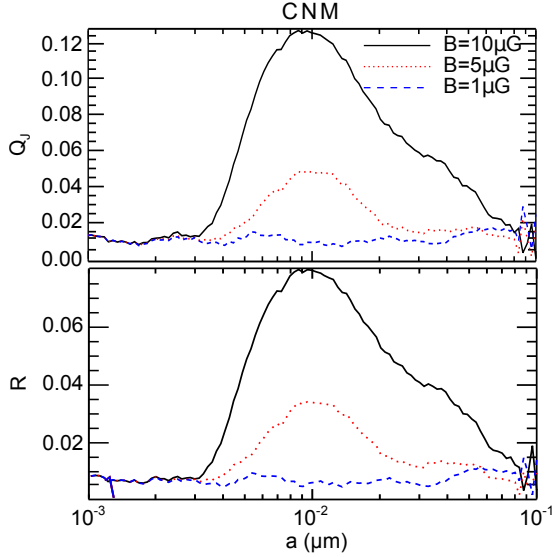}
\caption{Degrees of grain alignment by the D-G relaxation for thermally rotating grains with $K(\omega)$ changing with $\omega$. Upper and lower panels show $Q_{J}$ and $R$ as functions of $a$. Constant grain temperature $T_{\d}=20\K$ and three magnetic field strengths for the CNM are assumed. Results for silicate grains with axial ratio $r=2$ are shown.}
\label{fig:QJ_RL99_D-G}
\end{figure}

To clearly see the effect of fast rotation on the degree of paramagnetic alignment, we repeat calculations in the previous subsection using $K(\omega)$ from Equation (\ref{eq:Kappa_sil}). The obtained degrees of alignment $Q_{J}$ and $R$ are shown in Figure \ref{fig:QJ_RL99_D-G} for the CNM. As shown, both $Q_{J}$ and $R$ increase when $a$ decreases from $a=0.1\mum$ to $a\sim 0.01\mum$ during which the grain still rotates slowly and the paramagnetic relaxation rate increases. Below $a\sim 0.01\mum$, $Q_{J}$ and $R$ fall sharply as a result of the suppression of paramagnetic relaxation when the grain spins sufficiently fast, producing a peak alignment at this grain size.

\subsection{Resonance Paramagnetic Alignment of Subthermally Rotating Grains}\label{sec:alignment}
Below, we investigate the paramagnetic alignment by taking into account additional damping and excitation processes due to the collisions with ions, electric dipole emission, IR emission, and plasma drag. Due to these interaction processes, grains are expected to be rotating subthermally (i.e., $\omega< \omega_{th}$, see HDL10). We first consider the alignment by the D-G relaxation and then by both D-G relaxation and resonance relaxation. 

\subsubsection{Silicate Grains}

\begin{figure}
\centering
\includegraphics[width=0.4\textwidth]{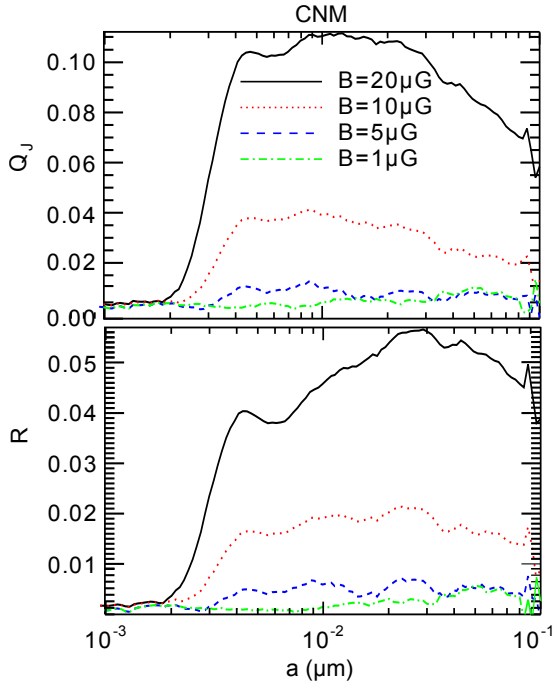}
\caption{Similar to Figure \ref{fig:QJ_RL99_D-G} but including various interaction processes (e.g., IR emission, plasma drag, and dipole emission). Upper and lower panels show $Q_{J}$ and $R$ as functions of grain size $a$.}
\label{fig:RQJ_D-G_CNM}
\end{figure}

Figure \ref{fig:RQJ_D-G_CNM} shows $Q_{J}$ and $R$ due to the D-G relaxation for silicate grains of axial ratio $r=2$. $Q_{J}$ and $R$ are shown in the upper and lower panels respectively. Similar to Figure \ref{fig:QJ_RL99_D-G}, one can see the sharp decline of $Q_{J}$ and $R$ at $a\sim 5\times 10^{-3}\mum$ as a result of the suppression of paramagnetic relaxation due to fast rotation. In particular, one can see a substantial decrease of grain alignment of $\sim 0.01\mum$ grains compared to Figure \ref{fig:QJ_RL99_D-G}. This is a direct consequence the additional damping processes included, which make grains to rotate subthermally and hence decrease the D-G alignment.

\begin{figure}
\centering
\includegraphics[width=0.4\textwidth]{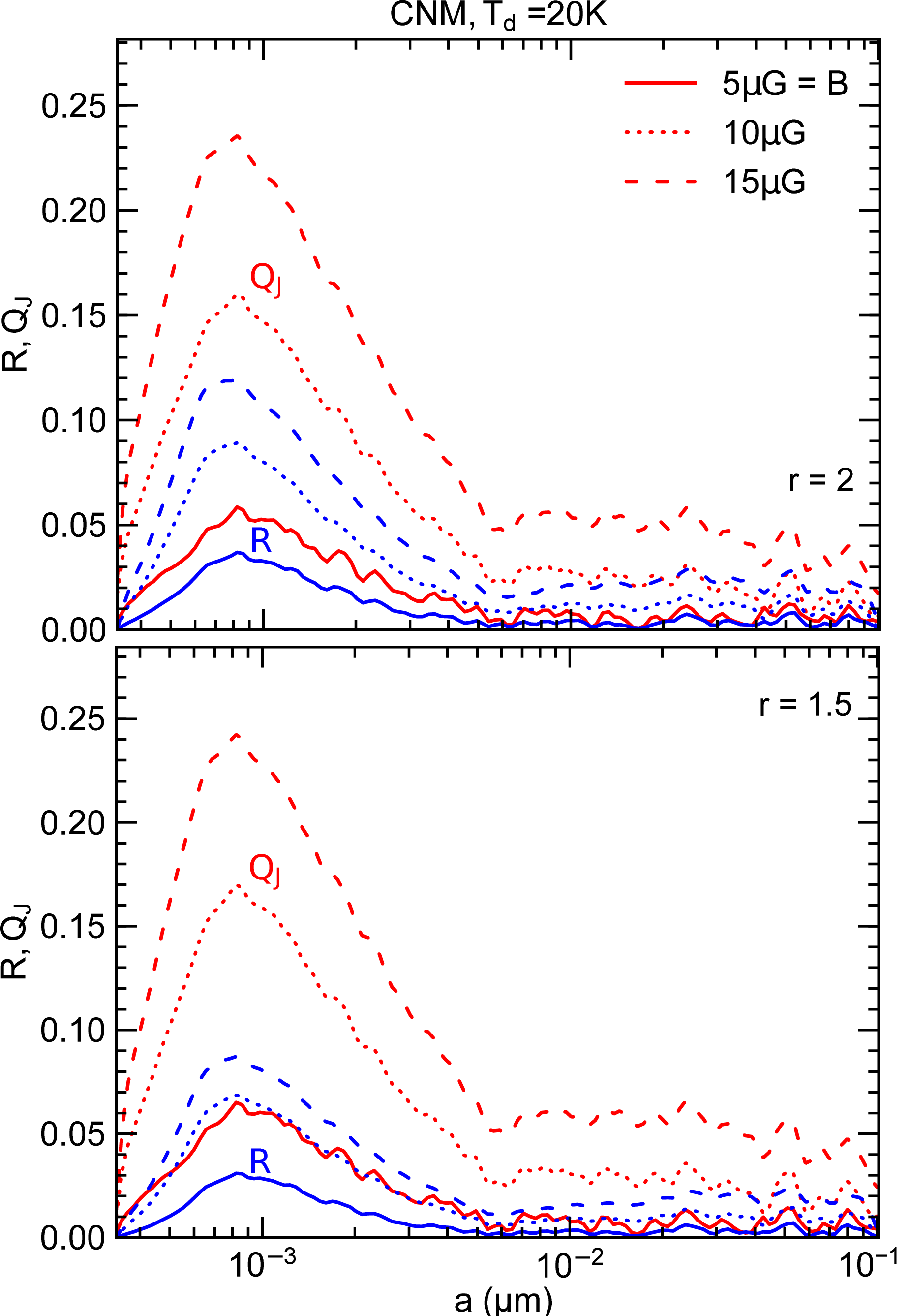}
\caption{Degrees of grain alignment by both the D-G relaxation and resonance relaxation for silicate grains with axial ratio $r=2$ (upper) and $r=1.5$ (lower). Grain temperature $T_{\d}=20\K$ is considered. Resonance relaxation induces the peaks of alignment around $10^{-3}\mum$.} 
\label{fig:RQJ_D-Gres_CNM}
\end{figure}

Figure \ref{fig:RQJ_D-Gres_CNM} shows $Q_{J}$ and $R$ as functions of grain size $a$ when the resonance relaxation is included for silicate grains with axial ratio $r=2$ (upper panel) and $r=1.5$ (lower panel). Compared to Figure \ref{fig:RQJ_D-G_CNM}, one can see that the resonance relaxation increases the alignment of ultrasmall grains, producing the peaks of alignment at $a\sim 10^{-3}\mum$. $Q_{J}$ is similar for two cases of grain shape while $R$ is smaller for the less elongated shape (lower panel) due to lower internal alignment. Some fluctuations in $Q_{J}$ and $R$ can be seen for $a>0.01\mum$ when they are as small as their numerical errors.

From the figure it can also be seen that the small ($\sim 0.01\mum$) grains are aligned much less efficient than the ultrasmall ($\sim 10^{-3}\mum$) grains if they have the same temperature $T_{\d}$. In fact, the temperature of ultrasmall grains is expected to be transient with temperature spikes due to UV heating, which decreases their alignment significantly. The temperature of the $a>0.01\mum$ grains in the ISM is estimated at $T_{\d}\approx 20\K$, thus from Figure \ref{fig:RQJ_D-Gres_CNM}, we can see that the paramagnetic alignment is rather small with $R<0.05$ for $B\le 15\mu$G. 
 
\begin{figure}
\centering
\includegraphics[width=0.4\textwidth]{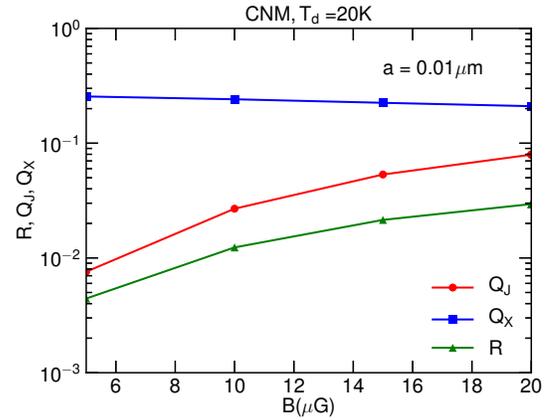}
\caption{Dependence of $Q_{X}, Q_{J}$ and $R$ as functions of the magnetic field strength for the $0.01\mum$ silicate grains of $T_{\d}=20\K$.}
\label{fig:RQJBfield}
\end{figure}

Figure \ref{fig:RQJBfield} shows the increase of $Q_{X}, Q_{J}$ and $R$ with $B$ for the $0.01\mum$ silicate grains of $T_{\d}=20\K$ in the CNM. As shown, $Q_{J}$ and $R$ increase rapidly with the increasing $B$ whereas $Q_{X}$ declines slowly with $B$.

\subsubsection{Carbonaceous Grains}

\begin{figure}
\centering
\includegraphics[width=0.4\textwidth]{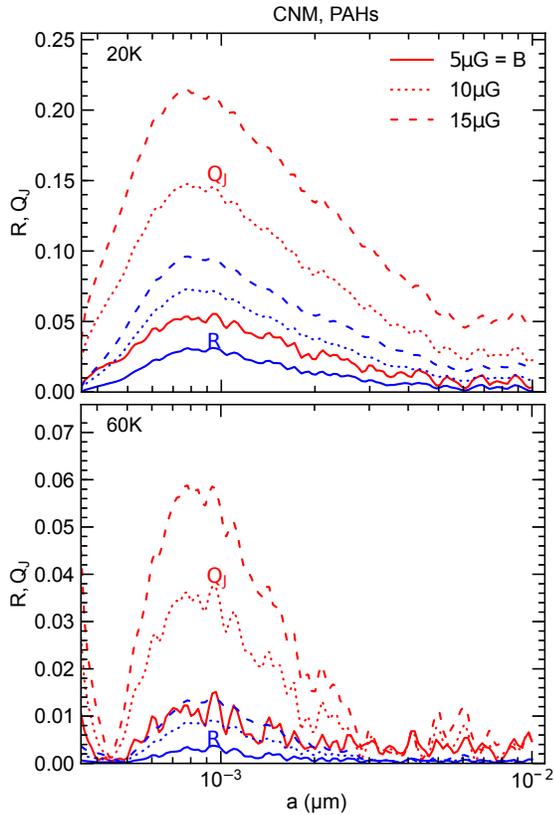}
\caption{$Q_{J}$ and $R$ of very small carbonaceous grains with axial ratio $r=2$ predicted for different $B$. Grain temperature $T_{\d}=20\K$ (upper) and $60\K$ (lower) are considered.}
\label{fig:RQJ_DGres_CNM_PAH}
\end{figure}

Figure \ref{fig:RQJ_DGres_CNM_PAH} shows $Q_{J}$ and $R$ computed for very small carbonaceous grains (i.e. PAHs) with the axial ratio $r=2$. Two grain temperature $T_{\d}=20\K$ and $60\K$ are considered. As shown, $Q_{J}$ and $R$ vary with the grain size $a$ in the same trend as silicate grains, although the degrees of alignment of PAHs are slightly lower than those predicted for silicate grains of the same $T_{\d}$ (see upper panel) due to the lower value of $f_{p}$. The degrees of alignment are subtantially decreased when the temperature is increased from $20\K$ to $60\K$.

One interesting feature for the higher $T_{\d}$ case is that $Q_{J}$ starts to rise at $a\approx 4$\AA. Such a feature may be caused by the excitation term of magnetic relaxation $B_{xx,\m}$. For large grains, the excitation by other processes dominate, but when $a\sim 4$\AA, this term dominates, resulting in the additional alignment.

For graphite grains, as discussed in the previous section, the paramagnetic alignment is expected to be negligible due to the rather low rate of paramagnetic relaxation.

\subsection{Paramagnetic alignment in the WIM}

\begin{figure}
\centering
\includegraphics[width=0.4\textwidth]{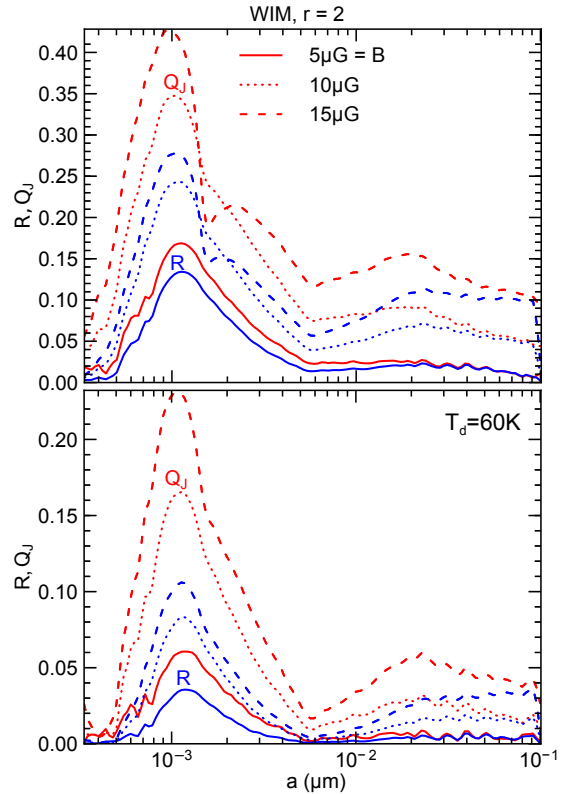}
\caption{Similar to Figure \ref{fig:RQJ_D-Gres_CNM} but for the WIM. The efficiency of magnetic alignment in the WIM is higher than in CNM due to its lower gas damping rate.}
\label{fig:RQJ_D-Gres_WIM}
\end{figure}

Figure \ref{fig:RQJ_D-Gres_WIM} shows the degrees of alignment of grains the WIM. It can be seen that the efficiency of paramagnetic alignment in the WIM is higher than in the CNM. Moreover, in contrast with the increase of $Q_{J}$ with the decreasing $a$ from $0.1-0.01\mum$ in the CNM, $Q_{J}$ decreases or almost is flat for $a$ in this range in the WIM. This is due to the fact that, as $a$ decreases, the ratio $\tau_{\gas}/\tau_{\DG}$ does not increase as in the CNM because in the WIM the dominant contribution to the rotational damping arises from IR emission, which has the timescale increasing with the decreasing grain size $a$ (see HDL10).

\section{Observational Constraints for Alignment of Small Grains}\label{sec:obs}
In this section, we are going to derive the grain size distribution and degree of grain alignment as a function of grain size (i.e., alignment function) that fit simultaneously to observed extinction and polarization curves. Let us first start with a summary on observational results for the starlight polarization.

\subsection{Observed Polarization Curves of Starlight}
Observational data in \cite{Serkowski:1975p6681} show that the polarization of starlight can be described well by an empirical law, usually referred to as the Serkowski law:
\bea
p(\lambda)=p_{\max}\exp\left[-K\ln^{2}\left(\frac{\lambda_{\max}}{\lambda}\right)\right],\label{eq:serkowski}
\ena
where $K$ is a parameter, which depends on $\lambda_{\max}$ (\citealt{1980ApJ...235..905W}). \cite{Whittet:1992p6073} derived the relationship $K=c_{1}\lambda_{\max}+c_{2}$ with $c_{1}=1.66\pm 0.09$ and $c_{2}=0.01\pm 0.05$ for most of sightlines. 

The observational data in \cite{Serkowski:1975p6681} also show that the maximum polarization of starlight is constrained by an upper limit 
\bea
p_{\max}\simlt 9\%E_{B-V},
\ena
which corresponds to 
\bea
p_{\max}\simlt 3\%A_{V}
\ena
for the typical diffuse ISM with $R_{V}=3.1$. For the general case, one expect that $p_{\max}/A(\lambda_{\max})\simlt 3\% {\rm mag}^{-1}$.

For some sightlines with low $\lambda_{\max}$ (e.g., $\lambda_{\max}<0.53\mum$), there exist an excess UV polarization from the Serkowski law (\citealt{1992ApJ...385L..53C}; \citealt{1995ApJ...445..947C}). The UV polarization for such sightlines can be described by a modified-Serkowski relation (\citealt{1999ApJ...510..905M}):
\bea
p_{\rm UV}=p_{\max}\exp\left[-K_{\rm UV}\ln^{2}\left(\frac{\lambda_{\max}}{\lambda}\right)\right],\label{eq:pUV}
\ena
where $K_{\rm UV}=\left(2.56\pm 0.38\right) \lambda_{\max}+\left(-0.59\pm 0.21\right)$.

In general, the variation of $p_{\max}/A(\lambda_{\max})$ from the upper limit can arise from fluctuations of the magnetic field direction from the perpendicular direction, the variation of the degree of grain alignment along the sightline, and the variation of grain properties (composition, shape). For instance, in molecular clouds, the decline of polarization efficiency $p_{\max}/A(\lambda_{\max})$ can be explained by the decline of the degree of grain alignment by radiative torques when going deeper into the cloud (\citealt{2005ApJ...631..361C}; \citealt{2008ApJ...674..304W}) or by the effect of magnetic turbulence \citep{1992ApJ...389..602J}. {\it The question is what is the imprint of the variation of the strength of magnetic fields on the polarization curves, provided that small grains are weakly aligned by paramagnetic relaxation?}

\subsection{Theoretical Considerations for Alignment Function}\label{sec:theor}

Recent advances in grain alignment theory allow us to predict the alignment of a variety of interstellar dust population, ranging from ultrasmall grains of a few Angstroms to micron-sized grains. As shown in Section \ref{sec:numeric}, ultrasmall and small grains can be aligned weakly by resonance paramagnetic and D-G paramagnetic relaxation while large grains are believed to be aligned efficiently by RATs. The grain size at which the RAT alignment starts to dominate is given by $a_{\ali}$, which is usually referred to as the critical size of aligned grains (see e.g., \citealt{2014MNRAS.438..680H}).

For the diffuse interstellar radiation field (ISRF, see \citealt{1983A&A...128..212M}), the value $a_{\ali}$ is determined by the maximum angular momentum induced by RATs, which is equal to (see \citealt{2008MNRAS.388..117H}; \citealt{2014MNRAS.438..680H}):
\bea
\frac{J_{\rm max}^{\rm RAT}}{J_{\rm th}}&=&\left(\int \Gamma_{\lambda} d\lambda\right) \frac{\tau_{\rm drag}}{J_{\th}},\\
&\approx &200\hat{\gamma}_{\rad}\hat{\rho}^{1/2}a_{-5}^{1/2}
\left(\frac{30\cm^{-3}}{n_{\H}}\right)\left(\frac{100\K}{T_{\gas}}\right)\nonumber\\
&&\times
\left(\frac{\bar{\lambda}}
{1.2\mum}\right)\left(\frac{u_{\rad}}{u_{\ISRF}}\right)\left(\frac{\overline{Q_{\Gamma}}}{10^{-2}}\right)
\left(\frac{1}{1+F_{\rm IR}}\right),~~~~~\label{eq:Jmax_RAT}
\ena
where $\tau_{\rm drag}=\tau_{\gas}/(1+F_{\rm IR})$ with $F_{\rm IR}$ being the damping coefficient due to IR emission, $\hat{\gamma}_{\rad}=\gamma_{\rad}/0.1$ with $\gamma_{\rad}$ the anisotropy degree of radiation field, and
\bea
\bar{\lambda}&=&\frac{\int \lambda u_{\lambda} d\lambda}{u_{\rad}},\label{eq:wavemean}\\
\overline{Q}_{\Gamma}&=&\frac{\int Q_{\Gamma}\lambda u_{\lambda}d\lambda}{\overline{\lambda}u_{\rad}},\label{eq:Qmean}
\ena
are the wavelength and RAT efficiency averaged over the entire radiation field spectrum, respectively. For grains of $a\ll \overline{\lambda}$ in the ISM, $\overline{Q}_{\Gamma}$ is approximately equal to
\bea
\overline{Q}_{\Gamma}\approx 2\left(\frac{\overline{\lambda}}{a}\right)^{-2.7}\approx 2.4\times 10^{-3}\left(\frac{\overline{\lambda}}{1.2\mum}\right)^{-2.7}a_{-5}^{2.7}.
\ena

For the ISRF of $\overline{\lambda}=1.2\mum$, the above equations yield a critical size (i.e., size for which $J_{\max}^{\rm RAT}=3J_{\th}$) of aligned grains $a_{\ali}\approx 0.05\mum$. As shown previously (e.g., \citealt{2005ApJ...631..361C}; \citealt{2009ApJ...695.1457H}), the value $a_{\ali}$ becomes larger for grains located deeper in molecular clouds (i.e., larger $A_{V}$). Thus, grains larger than $a_{\ali}$ are aligned efficiently by RATs while smaller grains ($a<a_{\ali}$) should be aligned weakly by the paramagnetic relaxation.

The degree of alignment $R$ of the $a>a_{\ali}$ grains tends to increase with increasing $a$ due to the increase of $J_{\max}^{\rm RAT}$ (i.e., less affected by randomization by gas bombardment). For small grains ($a<a_{\ali}$) that are being aligned by the paramagnetic relaxation, our computed results show that $R$ decreases with the decreasing $a$ (see Figures \ref{fig:RQJ_D-Gres_CNM}). The alignment of ultrasmall silicate grains ($a<5\times 10^{-3}\mum$) is peaky, but their contribution to the UV polarization is negligibly small. As a result, the alignment function of silicate grains that are important for producing the polarization curves is expected to increase with the increasing $a$.

\subsection{Observationally Inferred Grain Size Distributions and Alignment Functions}

To explore the variation of the alignment function $f(a)$ with $\lambda_{\max}$, we will find the best-fit models by fitting our theoretical models $p_{\mod}$ and $A_{\mod}$ (Equations \ref{eq:Plam} and \ref{eq:Aext}) to the observed polarization curves with $\lambda_{\max}=0.51\mum, 0.53\mum$ and $0.55\mum$. The observed polarization curves are calculated using Equations (\ref{eq:serkowski}) (for optical and IR wavelengths) and (\ref{eq:pUV}) (for UV wavelengths), taking the mean values of $K$ and $K_{\rm UV}$. The observed extinction curves are calculated using the extinction law (\citealt{1989ApJ...345..245C}; \citealt{1994ApJ...422..158O}) for $R_{V}= 3.1$. The search for best-fit models is performed by minimizing an objective function $\chi^{2}$ (see Appendix \ref{sec:chisq} for detail). We consider $N_{\lambda}=100$ bins of the wavelength from $\lambda=0.125-2.5\mum$ and $N_{a}=100$ bins of grain size from $a=3.56$\AA~to $1\mum$. We aim to perform the fitting for the case of maximum polarization efficiency, i.e., $p_{\max}/A(\lambda_{\max})=3\% {\rm mag}^{-1}$.

We adopt a mixture dust model consisting of amorphous silicate grains, graphite grains and PAHs (see \citealt{2001ApJ...548..296W} ; \citealt{2007ApJ...657..810D}). Since observational evidence for alignment of graphite is still missing, we conservatively assume that only silicate grains are aligned while carbonaceous grains are randomly oriented. Oblate spheroidal grains with axial ratio $r=2$ as in \cite{1995ApJ...444..293K} and $r=1.5$ are considered.

The fitting procedure is started with an initial size distribution $n(a)$ that best reproduces the observational data for the typical ISM, which corresponds to model 3 in \cite{Draine:2009p3780}. By doing so, we assume that dust properties are similar throughout the ISM and the difference in the polarization of starlight is mainly due to the efficiency of grain alignment, which depends on environment conditions along the sightlines, e.g., radiation field, magnetic fields and gas density. We take the alignment function for the ISM from \cite{Draine:2009p3780} as an initial alignment function. 

One particular constraint for the alignment function is that, for the maximum polarization efficiency $p_{\max}/A(\lambda_{\max})=3\% {\rm mag}^{-1}$, we expect that the conditions for alignment are optimal, which corresponds to the case in which the alignment of big grains can be perfect, and the magnetic field is regular and perpendicular to the sightline. Thus, we set $f(a=a_{\max})=1$. For a given sightline with lower $p_{\max}/A(\lambda_{\max})$, the constraint $f(a=a_{\max})$ should be adjusted such that $f(a=a_{\max})=(1/3)p_{\max}/A(\lambda_{\max})$. As discussed in Section \ref{sec:theor}, we expect the monotonic increase of $f(a)$ versus $a$, thus a constraint for this is introduced. Other constraints include the non-smoothness of $dn/da$ and $f(a)$ (see \citealt{2006ApJ...652.1318D}). 

The nonlinear least square fitting is carried out using the Monte Carlo direct search method. Basically, for each size bin, we generate $N_{\rm rand}$ random samples in the range $[-\zeta,\zeta]$ from a uniform distribution for $f(a)$ and $n(a)$, $\alpha_{f}$ and $\alpha_{n}$, respectively. The new values of $f$ and $n$ are given by $\tilde{f}=(\alpha_{f}+1)f(a)$ and $\tilde{n}=(\alpha_{n}+1)n(a)$. Then we calculate $p_{\mod}$ and $A_{\mod}$ for the new values $\tilde{f}$ and $\tilde{n}$ using Equations  (\ref{eq:Plam}) and (\ref{eq:Aext}). The values of $\chi^{2}$ obtained from Equation (\ref{eq:chisq}) are used to find the minimum $\chi^{2}$. The range $[-\zeta,\zeta]$ of the uniform distribution is adjusted after each iteration step. Initially $\zeta=0.5$ is assumed, which allows more room for the random sampling, and when the convergence is close (i.e., the variation of $\chi^{2}$ is small) $\zeta$ is decreased to $\zeta=0.1$.

The fitting procedure is repeated until convergence criterion is satisfied. Here, we use the convergence criterion, which is based on the decrease of $\chi^{2}$ after one step: $\epsilon=(\chi^{2}(n,f)-\chi^{2}(\tilde{n},\tilde{f}))/\chi^{2}(n,f)$. If $\epsilon\le \epsilon_{0}$ with $\epsilon_{0}$ sufficiently small, then the convergence is said to be achieved (see also \citealt{Hoang:2013dw}). With the value $\epsilon_{0}=10^{-3}$ adopted, the convergence is slow for some sightlines, then we stop the iteration process after 60 steps.

\begin{figure}
\centering
\includegraphics[width=0.4\textwidth]{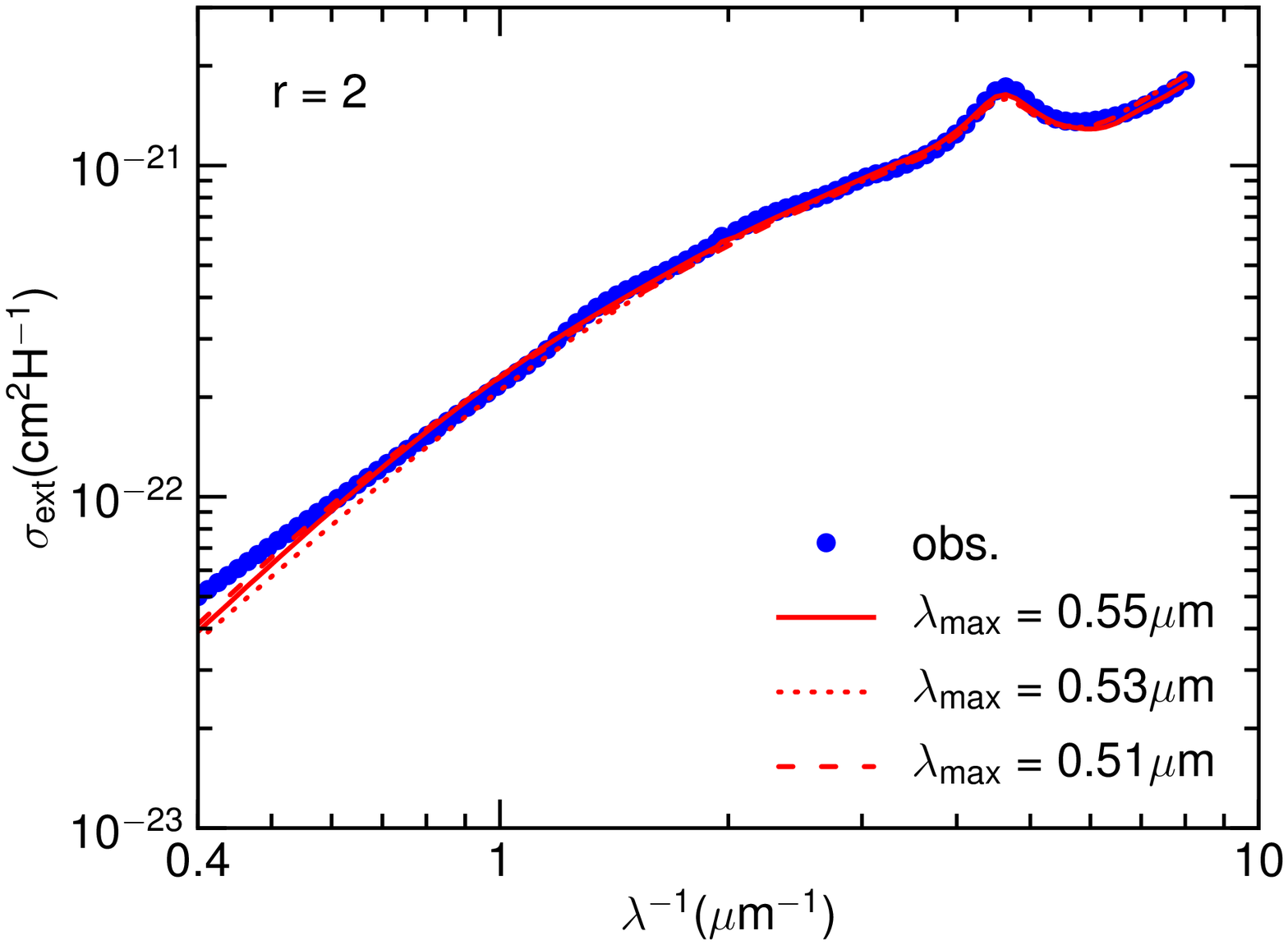}
\includegraphics[width=0.4\textwidth]{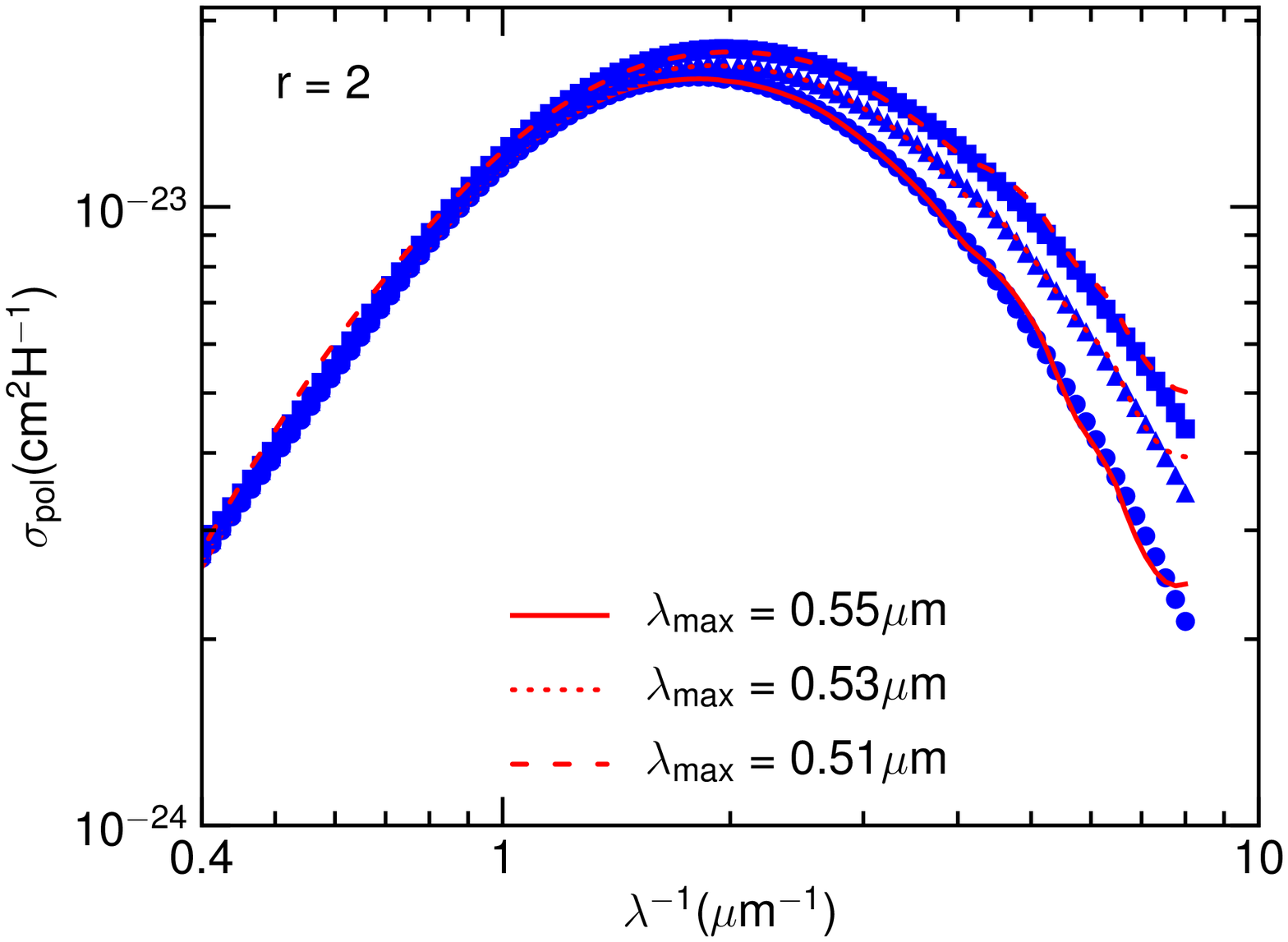}
\caption{Upper panel: observed extinction curve (symbols) of $R_{V}=3.1$ and best-fit models. Lower panel: observed polarization curves and best-fit models (solid, dotted and dashed lines). Three different values of $\lambda_{\max}$ and oblate grains with axial ratio $r=2$ are considered.}
\label{fig:bestfitmod}
\end{figure}

Figure \ref{fig:bestfitmod} (upper panel) shows the extinction cross section $\sigma_{\ext}$ as a function of $\lambda^{-1}$ for our best-fit models and the observed extinction curve with $R_{\V}=3.1$, assuming oblate spheroidal grains with axial ratio $r=2$. The lower panel shows $\sigma_{\pol}$ for our best-fit models and the observed polarization curves of different $\lambda_{\max}$. As shown, our models provide an excellent fit to the observational data in all cases of $\lambda_{\max}$. 

\begin{figure}
\centering
\includegraphics[width=0.4\textwidth]{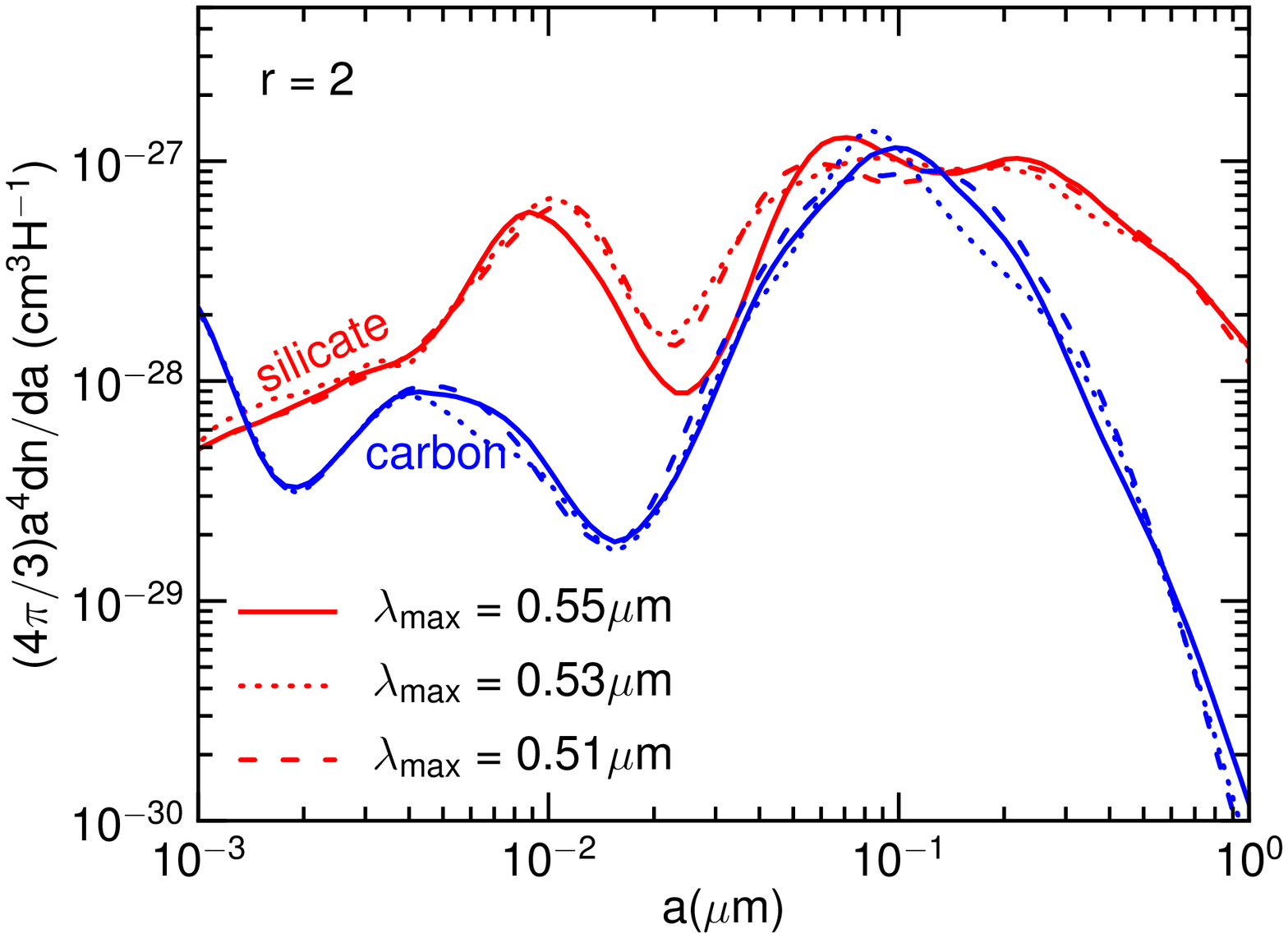}
\includegraphics[width=0.4\textwidth]{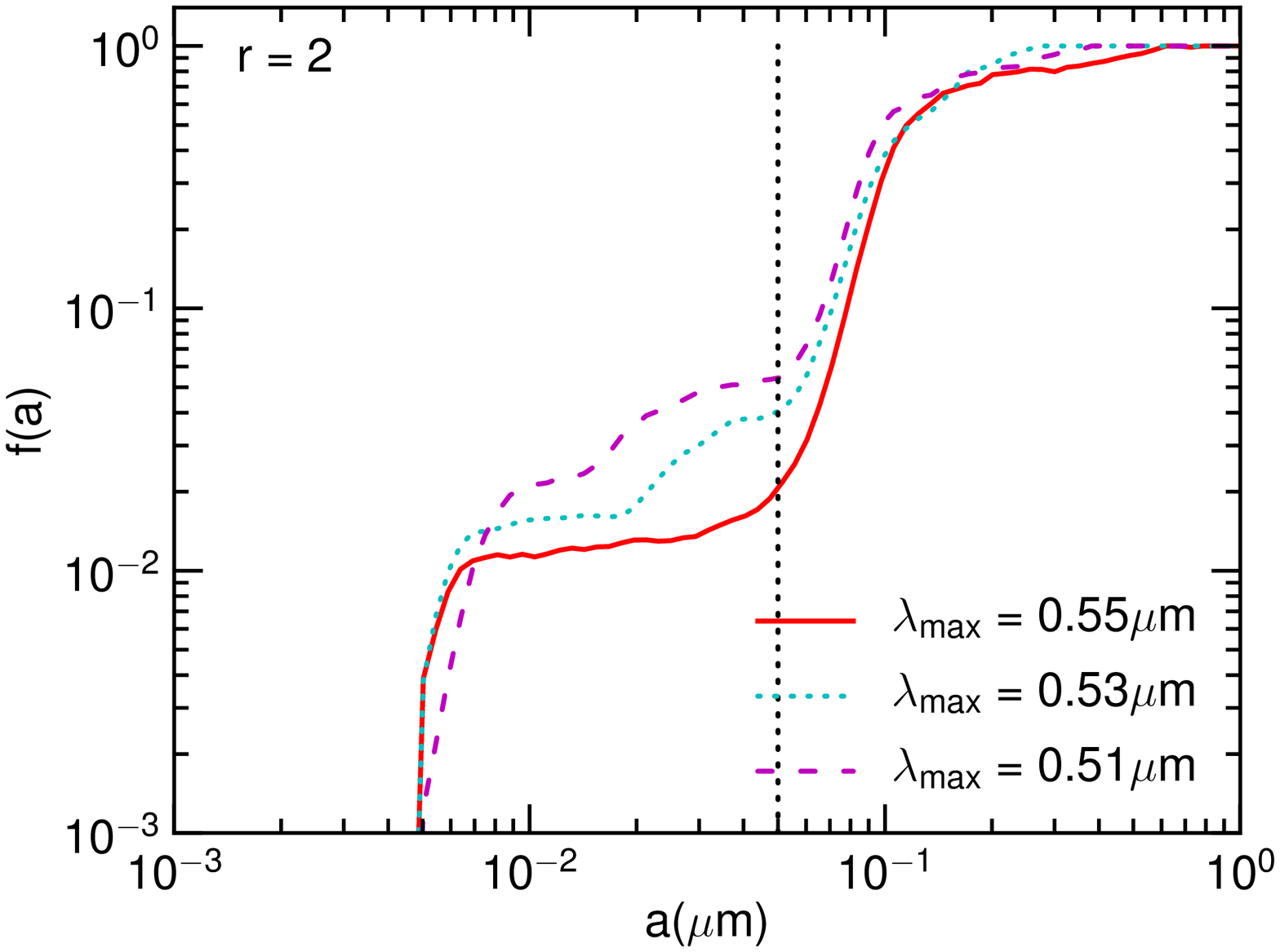}
\caption{Grain size distributions (upper panel) and alignment functions (lower panel) of our best-fit models for oblate grains with axial ratio $r=2$. The dotted vertical line marks $a=0.05\mum$. The size distribution appears quite similar for the different $\lambda_{\max}$, whereas the alignment of small grains ($a\sim 0.01-0.05\mum$) increases with the decreasing $\lambda_{\max}$.}\label{fig:fali}
\end{figure}

Figure \ref{fig:fali} (upper panel) shows the mass distributions $\propto a^{4}dn/da$ that reproduce the best-fit models in Figure \ref{fig:bestfitmod}. From the figure, one can see that our best-fit mass distributions of silicate grains have three peaks at $a\approx 0.01\mum, 0.07$ and $0.2\mum$. The mass of small grains in the range $a=0.01-0.05\mum$ is higher for lower $\lambda_{\max}$. 

Figure \ref{fig:fali} (lower panel) shows the alignment functions $f(a)$ for our best-fit models. One can see that the $a>0.1\mum$ grains are efficiently aligned with $f(a) > 0.5$ and then $f(a)$ drops rapidly for $a<0.1\mum$. Interestingly, a prominent transition from efficient alignment to weak alignment occurs at $a\sim 0.05-0.06\mum$ for all three cases of $\lambda_{\max}$, suggesting that this can be indicative of the change in the alignment mechanism (e.g., from RAT alignment to paramagnetic alignment). Moreover, the alignment degree of typical interstellar ($a>0.05\mum$) grains tends to shift to the range of smaller $a$ as $\lambda_{\max}$ decreases. In particular, as $\lambda_{\max}$ decreases, the degree of alignment of small grains $a\sim 0.01-0.05\mum$ must increase considerably in order to reproduce the observed polarization curves (see Figure \ref{fig:fali}, lower panel).

\begin{figure}
\centering
\includegraphics[width=0.4\textwidth]{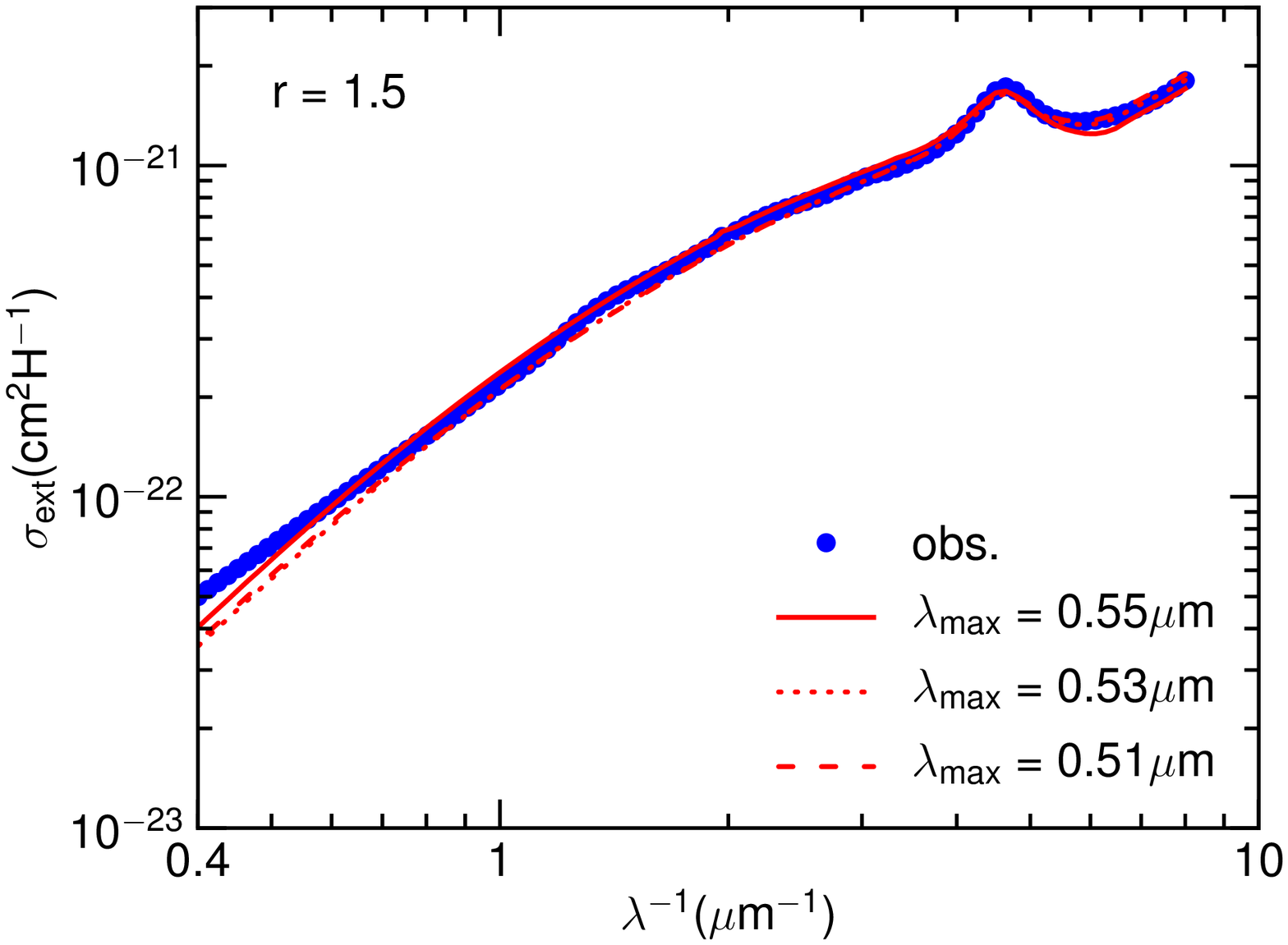}
\includegraphics[width=0.4\textwidth]{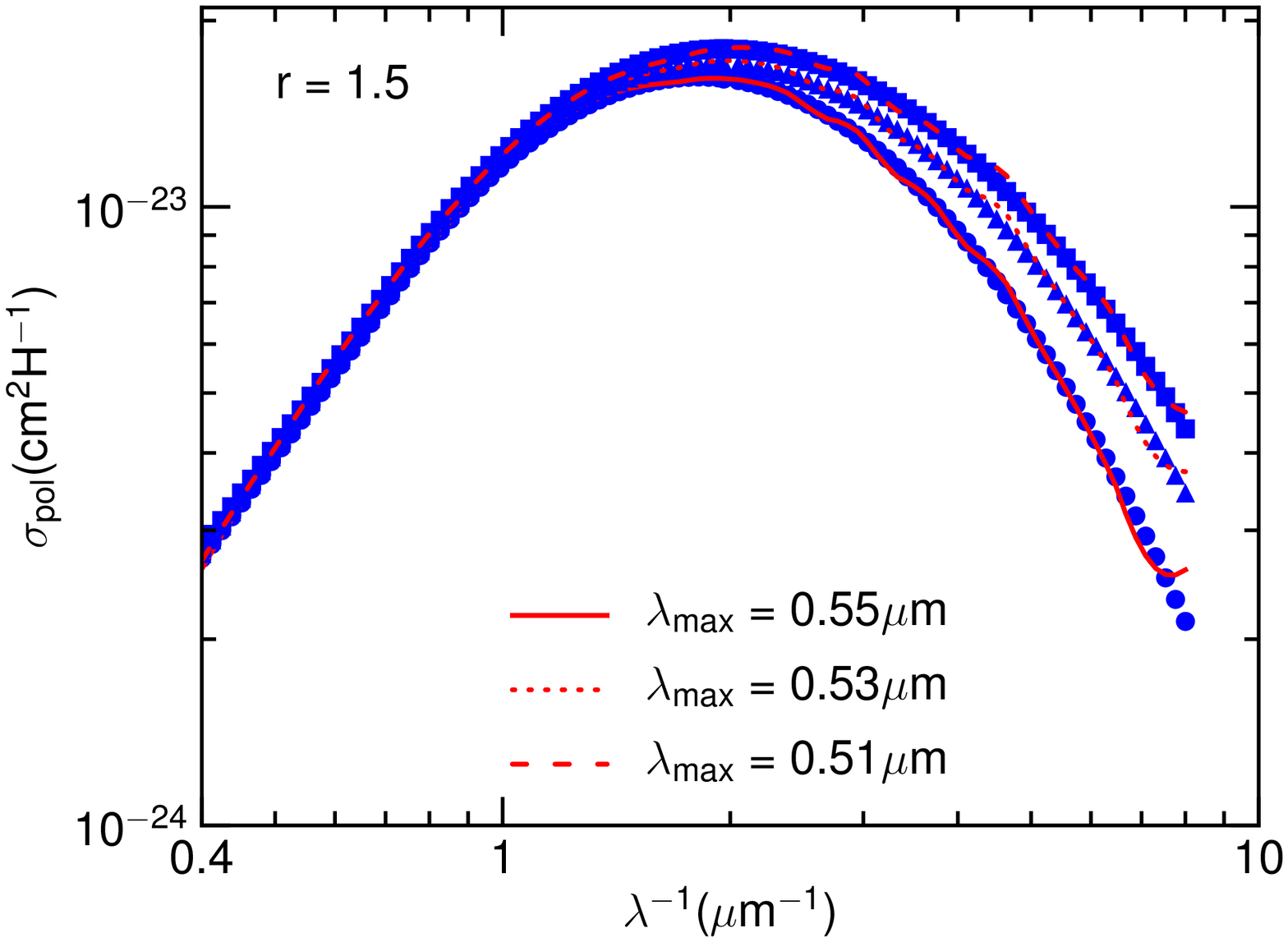}
\caption{Similar to Figure \ref{fig:bestfitmod} but for oblate grains of axial ratio $r=1.5$.}
\label{fig:bestfitmod-s067}
\end{figure}

\begin{figure}
\centering
\includegraphics[width=0.4\textwidth]{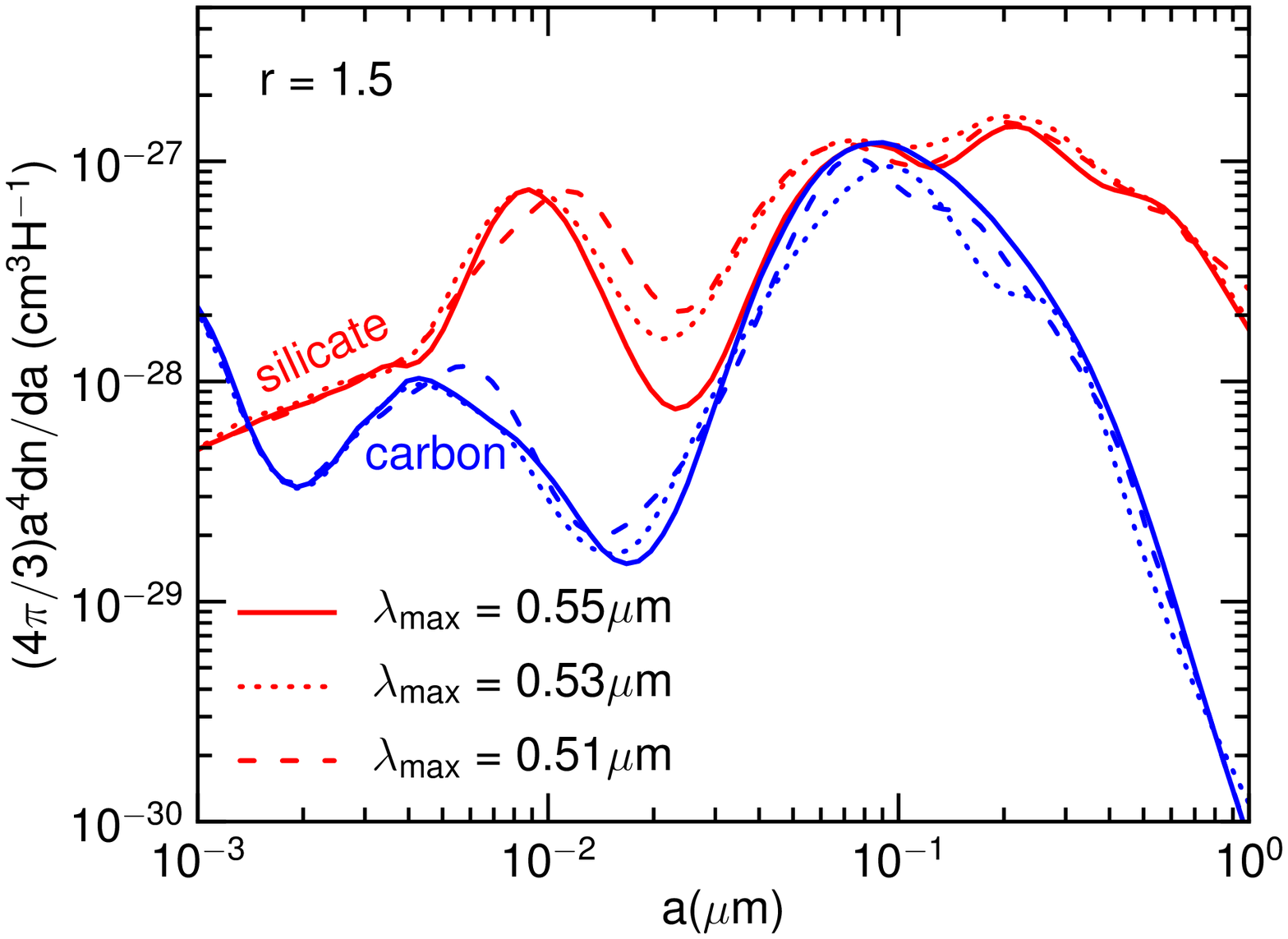}
\includegraphics[width=0.4\textwidth]{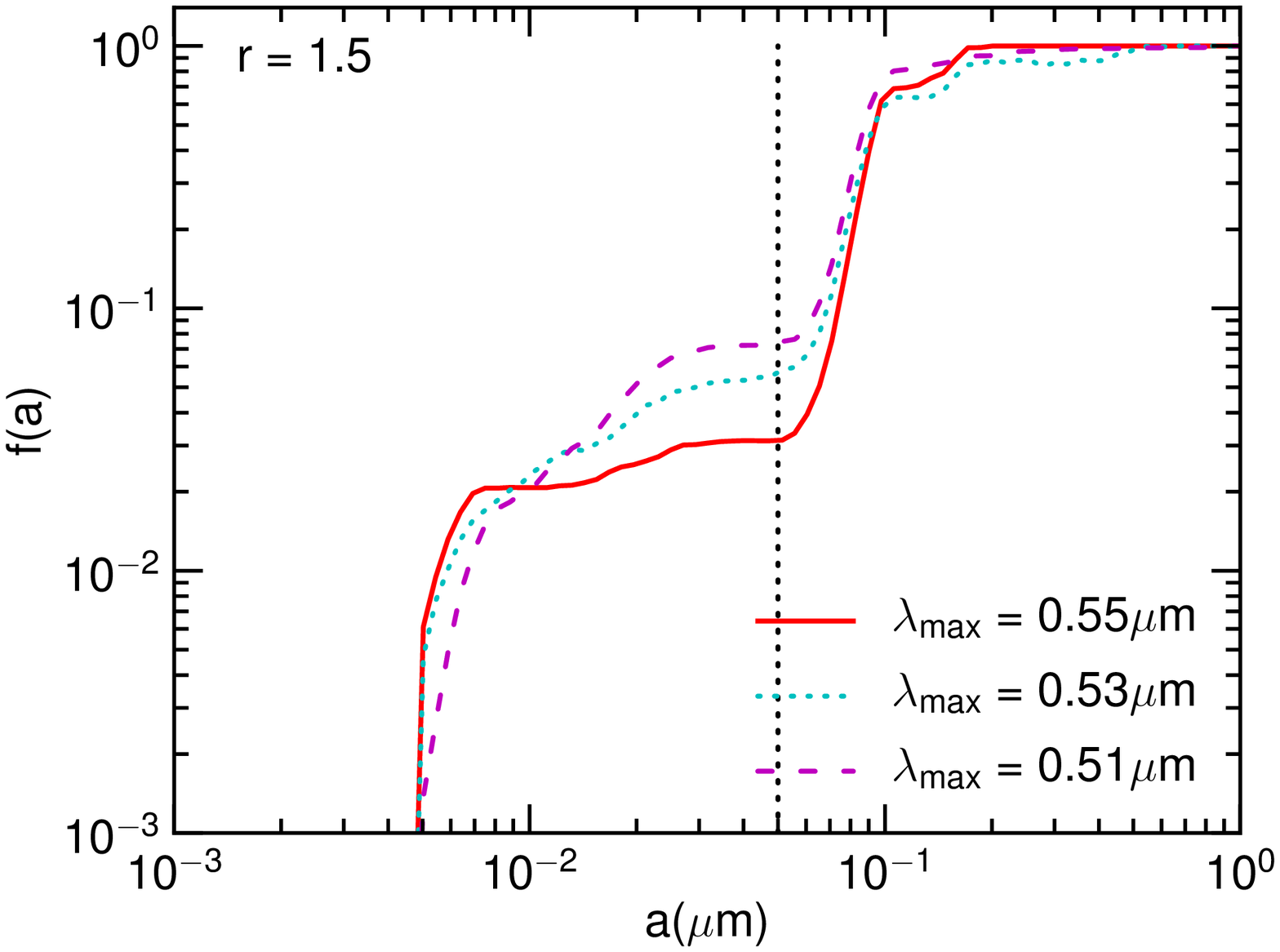}
\caption{Similar to Figure \ref{fig:fali} but for oblate grains with axial ratio $r=1.5$.}\label{fig:fali-s067}
\end{figure}

Similar to Figures \ref{fig:bestfitmod} and \ref{fig:fali}, Figures \ref{fig:bestfitmod-s067} and \ref{fig:fali-s067} show our best-fit models to the observed data for the case with axial ratio $r=1.5$. As shown, our models also provide good fit to the observational data. The alignment functions (see Figure \ref{fig:fali-s067}, lower) exhibit the same features (e.g., transition from efficient to weak alignment) as those in the case $r=2$. However, to reproduce the observed data, small grains with $r=1.5$ must have a degree of alignment higher than those with $r=2$ by a factor of $\sim 1.5$ (see the lower panels of Figures \ref{fig:fali} and \ref{fig:fali-s067}).

\section{Measuring Magnetic Fields using the UV polarization}\label{sec:Bfield}

In this section, we employ the degrees of alignment (from theoretical calculations and best-fit models) and size distributions obtained in the previous sections to predict theoretical polarization curves (see Section \ref{sec:extpol} for theory) for the diffuse ISM of the different magnetic field strengths.

\subsection{Theoretical Polarization Curves}
Since the fitting is performed for the case of maximum polarization efficiency $p_{\max}/A(\lambda_{\max})$ for which the magnetic field should lie in the sky plane, the inferred alignment function is then equal to the Raleigh reduction factor, i.e., $f(a)=R(a)$.

In the previous section, we found that the best-fit model requires the increased alignment of small grains as $\lambda_{\max}$ decreases. Such increased alignment of small grains in general can arise from (i) the increase of magnetic fields as calculated in Section \ref{sec:alignment} and (ii) the increase of RAT alignment due to enhanced radiation field by some hot stars in the vicinity of the sightline. In the latter case, the excess thermal emission is expected since dust is warmer due to higher radiation field. Below we consider the first situation and leave the second one for the discussion section.

To explore the effect of paramagnetic alignment of small grains on polarization curves, we distinguish the alignment of the typical interstellar grains with $a\ge a_{\ali}$ and that of smaller grains with $a< a_{\ali}$, which are expected to be induced by RATs and the paramagnetic relaxation, respectively. Moreover, there is always some intermediate range from the paramagnetic alignment to RAT alignment. Thus we assume that grains with $a\le a_{\rm mag}$ (i.e. $a_{\rm mag}<a_{\ali}$) are solely aligned by paramagnetic relaxation and take the degree of alignment computed in Section \ref{sec:numeric} for the CNM of different magnetic field strengths. The degree of alignment of grains with $a>a_{\rm mag}$ is taken from the best-fit alignment functions. The precise value of $a_{\rm mag}$ is uncertain, and we take $a_{\rm mag}\approx 0.04\mum$, which is equal to the grain size at which $J_{\max}^{\rm RAT}/J_{\th}=1$ for the diffuse ISM, i.e, when the RAT alignment is negligible. Moreover, since large grains are likely in thermal equilibrium with the ISRF while VSGs are expected to undergo thermal spikes due to the absorption of UV photons (\citealt{1989ApJ...345..230G}), we assume $T_{\d}=18\K$ for the $a>50$\AA~grains and $T_{\d}=60\K$ for very small ($a<50$\AA) grains.

\begin{figure*}[h]
\centering
\includegraphics[width=0.85\textwidth]{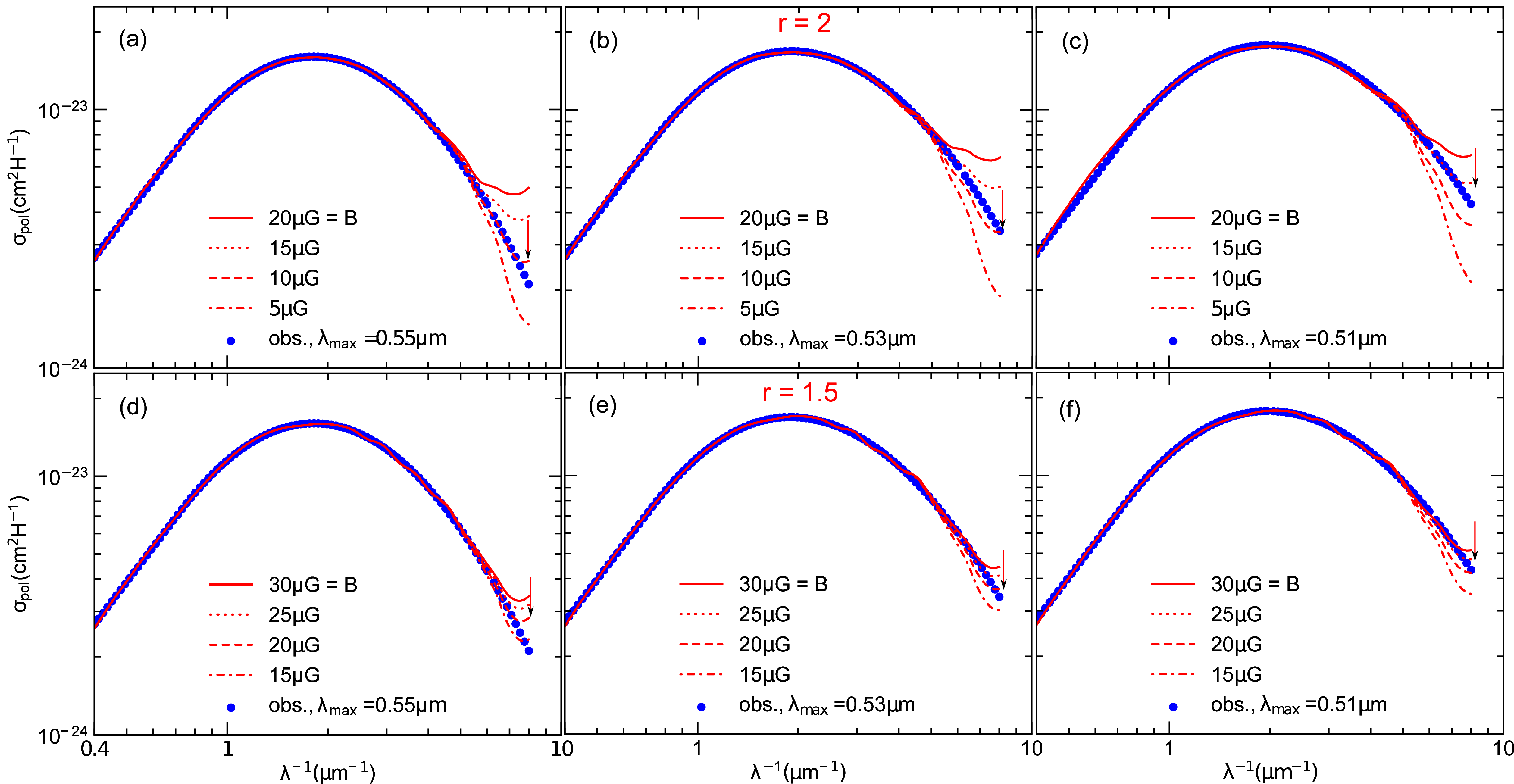}
\caption{Polarization cross-section arising from aligned silicate grains predicted for various values of the magnetic field $B$ versus the observed polarizations (filled circles) with different $\lambda_{\max}$. Upper and lower panels show results for oblate spheroids with axial ratio $r=2$ and $r=1.5$, respectively. The arrows indicate the theoretical curves that are close to the observational data.}
\label{fig:sigmapol_ISM}
\end{figure*}

Figure \ref{fig:sigmapol_ISM} shows $\sigma_{\pol}$ produced by aligned silicate grains for the different values $B$ for three selected $\lambda_{\max}$. Upper panels show results for the case axial ratio $r=2$ and lower panels show results for $r=1.5$. Filled circles show the observed polarization curves that are determined by $\lambda_{\max}$ (see Section \ref{sec:obs}). 

From the figure, we can see that the polarization at $\lambda^{-1}<5\mum^{-1}$ remains similar when changing $B$, indicating that the polarization at these wavelengths is determined by the alignment of typical interstellar grains ($a>0.05\mum$). On the other hand, the polarization in the UV ($\lambda^{-1}>5\mum^{-1}$) increases with the increasing magnetic field, which demonstrates that the alignment of the $a<0.05\mum$ grains by the paramagnetic relaxation plays an important role for the UV polarization. The rising feature of $\sigma_{\pol}$ computed at $\lambda^{-1}>7.5\mum^{-1}$ for some large $B$ arises from the fact that the best-fit alignment functions of small grains fall more rapidly with $a$ than computed theoretically assuming a constant $T_{\d}$.

For the case of $r=2$, the theoretical curve with $B=10\mu$G (dashed line, also indicated by the arrow) appears to be in good agreement with the observed curve of $\lambda_{\max}=0.55\mum$ (panel (a)). The corresponding values are $B \sim 15-20\mu$G for the cases with $\lambda_{\max}=0.53\mum$ and $0.51\mum$ (panels (b) and (c)). For the smaller axial ratio $r=1.5$, higher magnetic fields are required to reproduce the observed polarization curves in UV. For instance, $B\sim 15\mu$G for $\lambda_{\max}=0.55\mum$ (panel (d)) and $B \sim 20-25\mu$G for two other cases ((e) and (f)).

\subsection{Inferred Magnetic field Strengths}

As shown in the preceding subsection, higher magnetic fields are required to reproduce the observed UV polarization with lower $\lambda_{\max}$. To see clearly the dependence of the UV polarization on $B$ and $\lambda_{\max}$, we estimate the ratio $p(6\mum^{-1})/p_{\max}$ for the different $B$ and $\lambda_{\max}$. 

\begin{figure}
\centering
\includegraphics[width=0.40\textwidth]{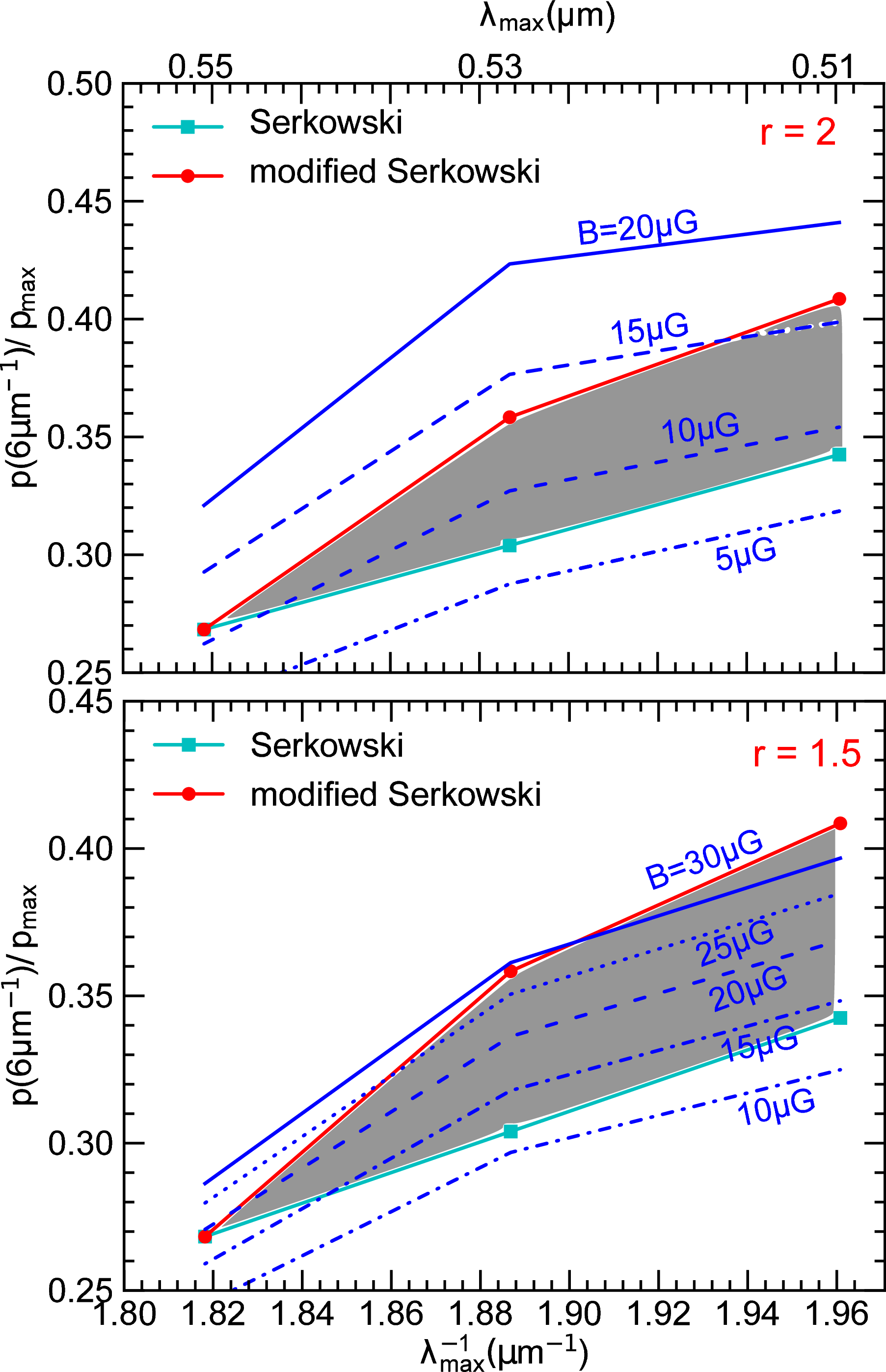}
\caption{Variation of $p(6\mum^{-1})/p_{\max}$ versus $\lambda_{\max}^{-1}$ predicted for different $B$ for spheroids with axial ratio $r=2$ (upper) and $r=1.5$ (lower). Observed data obtained from the Serkowski law (squares) and modified Serkowski law (circles) are also shown. Shaded areas show the intermediate range of $p(6\mum^{-1})/p_{\max}$ truncated by the predictions from the Serkowski and modified-Serkowski laws.}
\label{fig:pUV_contour}
\end{figure}

Figure \ref{fig:pUV_contour} shows $p(6\mum^{-1})/p_{\max}$ as a function of $\lambda_{\max}^{-1}$ predicted for the different values of $B$. Square and circle symbols show $p(6\mum^{-1})/p_{\max}$ calculated using Equations (\ref{eq:serkowski})(Serkowski law) and (\ref{eq:pUV}) (modified Serkowski law) with the mean values of $K$ and $K_{\rm UV}$. The magnetic field of the ISM seems to be well constrained in the range $B\sim 5-16\mu$G for $\lambda_{\max}=0.55-0.51\mum$ (upper panel) if the grain axial ratio $r=2$ is assumed. For less elongated spheroids of $r=1.5$, the range of magnetic field is $B\sim 10-30\mu$G for the same range of $\lambda_{\max}$ (lower panel). Specifically, for $\lambda_{\max}=0.55\mum$ the magnetic field strength is estimated at $B\sim 10\mu$G for axial ratio $r=2$, assuming the grain temperature $T_{\d}=18\K$. Estimated magnetic fields for $r=1.5$ are higher.

If the UV polarization is measured at $\lambda^{-1}>6\mum^{-1}$, the estimated magnetic field tends to be lower because the slope of $\sigma_{\pol}$ computed is shallower than the observed one (see arrows in Figure \ref{fig:sigmapol_ISM}).

\section{Discussion}\label{sec:discus}
\subsection{Comparison to previous studies on paramagnetic alignment}

The paramagnetic relaxation was introduced by \cite{1951ApJ...114..206D} to explain the alignment of interstellar grains with the Galactic magnetic field. The first quantitative study of grain alignment by paramagnetic relaxation (i.e., Davis-Greenstein (D-G) alignment) was carried out by \cite{Jones:1967p2924} using the Fokker-Planck (FP) equations. \cite{Purcell:1969p3641} and \cite{1971ApJ...167...31P} studied the D-G alignment by means of the Monte Carlo method and showed that this mechanism is inefficient in aligning the typical interstellar grains. Latter, \cite{1979ApJ...231..404P} suggested that the joint action of spin-up systematic (pinwheel) torques that can drive grains to suprathermal rotation and the paramagnetic relaxation could result in efficient alignment of suprathermally rotating grains. However, the efficiency of pinwheel torques is believed to be significantly suppressed due to the rapid thermal flipping of small grains, and small grains are expected to be thermally trapped (\citealt{1999ApJ...516L..37L}; \citealt{2009ApJ...695.1457H}). Therefore, we disregarded minor effects of pinwheel torques on the alignment of small grains. 

For thermally rotating grains, \cite{Lazarian:1997p5348} calculated the paramagnetic alignment using an analytical method based on the FP equations. RL99 have computed the efficiency of the D-G mechanism for these grains using the Langevin equations. Both papers took into account the Barnett relaxation effect and internal thermal fluctuations (an inverse process associated with the Barnett relaxation). Nevertheless, these aforementioned studies assumed a constant magnetic susceptibility $K(\omega)$ and considered the rotational damping and excitation due to dust-gas collisions only. Such assumptions are obviously valid for large ($>0.1\mum$) grains that rotate slowly in the absence of pinwheel torques. Their essential conclusion is that the D-G mechanism is inefficient in aligning the interstellar grains and failed to account for the observed polarization in the molecular clouds where the dust and gas are likely in thermal equilibrium.

This paper investigates the paramagnetic alignment for a wide range of grains, from a few Angstroms to $0.1\mum$, using the Langevin equations (RL99; HDL10). This grain population is expected to rotate subthermally but rapidly with $\omega>10^{5}\s^{-1}$. We take into account the various damping and excitation processes that are essential for the rotational dynamics of small grains, including gas-dust collisions, plasma drag, IR emission, and electric dipole damping (see \citealt{1998ApJ...508..157D}; \citealt{2010ApJ...715.1462H}). For small grains, we found that the efficiency of paramagnetic alignment is indeed rather low due to subthermal rotation; the degree of paramagnetic alignment $R$ increases with the magnetic field strength $B$, but $R < 0.05$ for $B<20\mu$G for the typical ISM conditions.

\cite{2000ApJ...536L..15L} (LD00) have identified a new physical process, namely, resonance paramagnetic relaxation, which is shown to enhance the alignment of ultrasmall grains. The efficiency of resonance paramagnetic alignment was estimated at a level of $10\%$ for the $10$\AA~grains in LD00 where the idealized model of spinning dust emission from \cite{1998ApJ...508..157D} (DL98) was adopted. The present work used an improved model of spinning dust emission from HDL10 which accounts for the grain wobbling and quantified the efficiency of grain alignment by resonance paramagnetic relaxation. Our results in general confirmed the predictions by \cite{2000ApJ...536L..15L}. The only difference is that our results predict a lower grain size (about 10\AA) of the peak alignment than earlier predicted by LD00. This difference arises from the improved model of spinning dust that predicts lower rms grain angular momentum than the DL98 model.

\subsection{Excess UV polarization and Alignment of Small Grains}

The excess of continuum polarization in the UV with respect to the Serkowski law, usually characterized by $p(6\mum^{-1})/p_{\max}$, was observationally reported in \cite{1992ApJ...385L..53C} and \cite{1995ApJ...445..947C} (see also \citealt{1999ApJ...510..905M}). However, it is still unclear why such an excess UV polarization only exists for $\lambda_{\max}<0.55\mum$. {\it To resolve this question, we first need to understand which grain population is responsible for the UV polarization.}

The original Serkowski law fits well to the observed polarization at IR and optical wavelengths. At these wavelengths, we showed that the polarization is mostly produced by typical silicate grains ($a> 0.05\mum$) aligned in the magnetic field (see Section \ref{sec:Bfield}). However, the polarization in the UV arising from these relatively large grains is insufficient to reproduce the observed polarization; the contribution of weakly aligned small silicate grains ($a<0.05\mum$) allows us to successfully reproduce the UV polarization.

If the excess UV polarization is indeed produced by small aligned grains, then why the alignment of this grain population increases, as it is required by higher $p(6\mum^{-1})/p_{\max}$, in the cases $\lambda_{\max}<0.55\mum$?

Compared to the typical polarization curve of the ISM with $\lambda_{\max}=0.55\mum$, we found that, for the cases with $\lambda_{\max}<0.55\mum$, the alignment function of grains tends to shift to the smaller grain size, corresponding to the decrease of critical size of aligned grains $a_{\ali}$. Thus, there exists some additional alignment of {\it intermediate size} grains ($a=0.05-0.1\mum$) by the same alignment mechanism as typical interstellar grains (most likely driven by RATs), which gives rise to shift the polarization curve to the shorter $\lambda$. At the same time, we found that the alignment of small grains $a<0.05\mum$ must be enhanced to reproduce the excess UV polarization. Thus, there seems to exist some correlation between the alignment of typical interstellar grains, which is most likely driven by RATs, and the alignment of small grains. Below, we discuss some possible reasons why this could happen.

If the enhanced alignment of small grains is induced by increased RATs due to nearby hot stars, then such a correlation is obvious. However, some stars that have the excess UV polarization do not exhibit excess thermal emission at $60\mum$ (see \citealt{1995ApJ...445..947C}). Interesting enough, the HD197770 star possesses an excess emission at $60\mum$, but has actually a lower excess UV polarization (see \citealt{1993PASP..105.1127G}; \citealt{1995ApJ...445..947C}). This indicates that dust along these sightlines with the excess UV polarization is actually not hotter than the dust along the stars without the excess. In addition, the amount of dust near the stars may be rather small compared to the total dust mass along the entire sightline, as suggested in \citep{1995ApJ...445..947C}.  Furthermore, if the enhanced alignment of small grains is caused by RATs, then the sharp transition in the alignment function at $a\sim 0.05\mum$ for the best-fit models is unexpected because $f(a)$ should decrease monotonically from $a=0.1\mum$ to $a\sim 0.03-0.04\mum$ as seen in the alignment function obtained for HD 197770. Therefore, the enhanced alignment of small grains by increased RATs may not be a dominant reason for the excess UV polarization.

If the enhanced alignment of small grains is induced by an increased magnetic field strength, then the correlation can be due to the following reasons.

First,  the RAT alignment tends to increase with increasing magnetic field strength as paramagnetic alignment. Indeed, in the RAT alignment paradigm, we find that the increase of the paramagnetic relaxation can result in the increase of the fraction of grains aligned with high-$J$ attractor points, which increases the degree of RAT alignment (\citealt{2007MNRAS.378..910L}; \citealt{2008MNRAS.388..117H}; \citealt{2008ApJ...676L..25L}). 

Second, the grain randomization due to the electric field acting on the electric dipole moment of grains that are accelerated by interstellar turbulence (\citealt{2002ApJ...566L.105L}; \citealt{2003ApJ...592L..33Y}; \citealt{2004ApJ...616..895Y}; \citealt{Yan:2009p5765}; \citealt{2012ApJ...747...54H}) is found to decrease (i.e., the degree of RAT alignment is increased) when the magnetic field is increased. The effect of such a randomization is described in \cite{2006ApJ...647..390W} and \citep{2009MNRAS.400..536J}.\footnote{We disagree with the conclusions of these studies, but accept the existence and potential importance of the randomization.} For a weak magnetic field, the randomization is thought to be more important because the rate of Larmor precession is lower than the rate of dipole fluctuations. As the magnetic field increases, the RAT alignment is expected to increase because the Larmor precession frequency becomes larger, reducing the randomization effect by dipole fluctuations.

\subsection{Measuring Magnetic Fields using the UV Polarization}

Magnetic fields are no doubt important for numerous astrophysical processes, including star formation, transport and acceleration of cosmic rays, and accretion disks. Dust polarimetry proves being a useful technique to trace the magnetic field direction in molecular clouds, and when combined with the Chandrasekhar-Fermi (CF) technique (\citealt{1953ApJ...118..113C}) one can measure the magnetic field strength.

While the variation of the local magnetic field direction along a sightline is usually referred to explain why the observed $p_{\max}/A(\lambda_{\max})$ is lower than its upper limit $p_{\max}/A(\lambda_{\max})= 3\% {\rm mag}^{-1}$, the effect of the magnetic field strength on the polarization curve has not been explored yet. The present study showed that the magnetic field strength can have important imprints on the observed polarization curves, particularly, it results in the excess UV polarization for cases $\lambda_{\max}<0.55\mum$. Using this subtle effect, we can estimate the strength of interstellar magnetic fields.

Assuming the average ISRF and grain axial ratio $r=2$, we find that, for the typical diffuse ISM with $\lambda_{\max}=0.55\mum$, the magnetic field strength is estimated at $B\sim 10\mu$G. This magnetic strength appears to be consistent with the Zeeman measurements (see \citealt{2012ARA&A..50...29C} for a recent review). For the sightline with $\lambda_{\max}=0.53\mum$ and $\lambda_{\max}=0.51\mum$, the estimated magnetic fields are $B\sim 13\mu$G  and $B=16\mu$G (see Figure \ref{fig:pUV_contour}, upper). Therefore, the magnetic field tends to increase with the decreasing $\lambda_{\max}$. When the grain axial ratio $r=1.5$ is considered, then the magnetic fields estimated for the selected sightlines would be higher.

Our above estimates for the magnetic field strength were carried out for the three idealized sightlines that have the optimal conditions for grain alignment, e.g., perpendicular magnetic field and perfect alignment of biggest grains. Therefore, the estimated magnetic fields correspond to the upper limits of the magnetic fields. 

Moreover, the diffuse ISM is known to be turbulent, which is a leading cause for the variation of $p_{\max}/A(\lambda_{\max})$ for different sightlines (see \citealt{2014arXiv1405.0871P}). For some sightline having $p_{\max}/A(\lambda_{\max})<3\% {\rm mag}^{-1}$ but the same $\lambda_{\max}$ and $p(6\mum^{-1})/p_{\max}$ as our selected sightlines (i.e., $p_{\max}/A(\lambda_{\max})=3\% {\rm mag}^{-1}$), the magnetic field strength would be similar to that with the maximum $p_{\max}/A(\lambda_{\max})$ if we assume the increase of $p_{\max}/A(\lambda_{\max})$ is due to the fluctuation of $\Bv$ and that the biggest grains can still be perfectly aligned. The reason for that is that the strength of $\Bv$ depends on the Rayleigh reduction factor $R$, which is the same in two sightlines while the effective degree of alignment $f$ changes as $f=R\cos^2\xi$. If both the fluctuations of $\Bv$ and unfavorable conditions of grain alignment responsible for lower $p_{\max}/A(\lambda_{\max})$, then the magnetic fields should be lower than the magnetic fields estimated for the idealized sightlines.

One of the important implications of this study is that it provides us a novel way to measure the strength of the magnetic field vector using three observational polarization parameters $p_{\max}, \lambda_{\max}$ and $p(\rm UV)$.  This technique is more useful for the sight lines with low $\lambda_{\max}$ because the UV polarization is not too low compared to the $p_{\max}$. The presented method allows us to obtain a constraint on the strength of the total magnetic field, which is more advantageous than other methods that return the projected magnetic field only. { It is also worth to mention that the usage of $\lambda_{\max}$ as an input parameter for measuring $B$ is cautious because of its complicated dependence on other parameters, including $R_{V}$ and $A_{V}$ (see \citealt{2007ApJ...665..369A}).}

{ The present method for measuring magnetic fields makes use of the polarization data in UV wavelength range from the Wisconsin Ultraviolet Photo-Polarimeter Experiment (WUPPE), which is below the atmospheric cut-off ($\sim 0.3\mum$). Therefore, to apply this method for beyond WUPPE data set, new space/rocket missions would be needed.} 

\subsection{Dependence of Inferred Magnetic fields on physical parameters}

There exists a number of parameters that appear to affect the inferred magnetic field strength using the UV polarization. 

First, grain geometry (i.e., asphericity) can affect the inferred magnetic fields. Our study considered two cases of oblate spheroidal grains with axial ratio $r=2$ and $r=1.5$. The latter grain shape has lower polarization cross-section $C_{\pol}$, and the degree of alignment required to reproduce the observational data is higher, resulting in the stronger inferred magnetic fields.  

Second, the grain temperature of small grains may also play an important role on the estimated magnetic field. Because the temperature of small dust grains determines the level of thermal fluctuations of grains axes with its angular momentum, which constrains the degree of internal alignment, our estimated magnetic field strengths based on the UV polarization should vary with the dust temperature chosen. Nevertheless, the temperature of small $\sim 0.01\mum$ grains is expected to be nearly stable in thermal equilibrium (see \citealt{2003ARA&A..41..241D}), so we expect the effect of grain temperature fluctuations plays a minor role for constraints of $B$ field.  

Third, the alignment of small grains is completely attributed to the paramagnetic alignment. Indeed, the alignment may be enhanced due to the additional effect of pinwheel torques (e.g., H$_{2}$ formation, see \citealt{2013ApJ...775...84A}). 

Fourth, our finding that the magnetic field tends to increase with the decreasing $\lambda_{\max}$ is based on the assumption that the average ISRF (e.g., $a_{\ali}$) is similar along the three sightlines. This assumption is valid for most of the sightlines with the excess UV polarization but do not exhibit excess thermal emission. For some sightlines with both the excess UV polarization and thermal emission, the magnetic field required to reproduce the observed polarization may not need to be increased.

Finally, when the strength of magnetic field is known, we can constrain the grain physical properties, such as grain geometry, using the UV polarization. Earlier studies (\citealt{1995ApJ...444..293K}; \citealt{2006ApJ...652.1318D}; \citealt{Draine:2009p3780}) and our present work show that a wide range of axial ratio of oblate spheroid can reproduce the observed extinction and polarization curves. However, grains with a small/large axial ratio (i.e. less/more elongated) will require a higher/lower degree of alignment of small grains, which corresponds to higher/lower magnetic fields, to reproduce the observed polarization. Thus, it is potential to constrain the grain geometry when the magnetic fields are known.

\subsection{Resonance Paramagnetic Alignment of ultrasmall Grains and Polarization of Spinning Dust Emission}

\cite{Hoang:2013dw} showed that the 2175\AA~polarization bump of HD 197770 can be reproduced successfully by a model of aligned silicate plus weakly aligned PAHs. The alignment function for their best-fit model has peak of $R\cos^2{\xi}\approx 0.004$ at $\sim 10$\AA. Accounting for the possible magnetic field orientation, assuming that the magnetic orientation results in $p_{\max}/A_{V}(\lambda_{\max})<3\% \mag$ (the upper limit of $p_{\max}/A_{V}(\lambda_{\max})$ of this star, one obtain $R\approx 0.006$.

We computed exactly the degree of alignment for VSGs (e.g., PAHs) for different magnetic field and temperature. For the case $T_{\d}=60\K$, we found the peak alignment $R\sim 0.006$ for $B \sim 5\mu$G (see Figure \ref{fig:RQJ_DGres_CNM_PAH}), which is equal to the alignment degree for the best-fit model in \cite{Hoang:2013dw}.

The question is why only HD 197770 posses the $2175$\AA~polarization bump but other stars with the similar $\lambda_{\max}$ do not?

It is noted that the possibility to observe the $2175$\AA~ polarization bump depends on both the alignment of PAHs and small silicate grains because the latter is responsible for the UV continuum polarization at $\lambda^{-1}>5\mum^{-1}$. If the alignment of small silicates is inefficient, then the bump can be detected due to high contrast. If the alignment of small silicate grains is considerable, the UV polarization produced by such grains tends to smooth out the bumpy polarization by PAHs, which makes the detection of 2175\AA~bump more difficult.

One interesting point in the polarization curve of HD 197770 is that its excess UV polarization is much lower than other stars with the same $\lambda_{\max}=0.51\mum$ (see \citealt{1995ApJ...445..947C}). On the other hand, the HD 197770 has an excess emission at $60\mum$, indicating that the radiation field is higher than the averaged ISRF and the dust is hotter than the typical ISM. Since hotter dust tends to reduce the alignment of small grains, the UV continuum polarization is reduced as well, favoring the detection of the 2175\AA~polarization bump. 

{ A related issue is the alignment of carbonaceous grains and its consequence. PAHs are thought to have attachment of aliphatic structures to its surface, producing large carbonaceous grains \citep{2011Natur.479...80K}. However, the idea that PAHs can be weakly aligned by resonance relaxation seems not to contradict with the unpolarized $3.4\mum$ aliphatic features \citep{2006ApJ...651..268C}. Indeed, if there is attachment of aliphatic structures to a PAH, the net size of aliphatic-PAH grain will increase, which makes the grain to rotate slower, assuming the same gas temperature. As a result, the alignment of the aliphatic-PAH grain by resonance relaxation would become negligible. The alignment of large carbonaceous grains by radiative torque may also be inefficient as discussed in a recent review by Lazarian et al. (2014).}

\subsection{Relating the UV polarization of starlight to spinning dust polarization}
Based on the UV polarization of starlight, one can infer the degree of alignment of small grains. Since the alignment of small grains and ultrasmall grains is most likely induced by the same paramagnetic mechanism, we can derive the alignment of ultrasmall grains. Then, the polarization of spinning dust can be constrained using the inferred degree of alignment of VSGs (see \citealt{Hoang:2013dw}).

\section{Summary}\label{sec:sum}
We calculated the degree of grain alignment by the Davis-Greenstein relaxation and resonance paramagnetic relaxation for subthermally rotating grains, and suggested a new way to constrain magnetic field strength using UV polarimetry. Our principal results can be summarized as follows.

\begin{itemize}

\item[1.] The degrees of grain alignment by paramagnetic relaxation (classical Davis-Greenstein and resonance one) were calculated for both small grains ($a\sim 0.01\mum$) and ultrasmall grains ($a\sim 0.001\mum$). We found that the alignment of small grains is dominated by the D-G relaxation while the alignment of ultrasmall grains is dominated by the resonance relaxation. The degree of alignment for normal paramagnetic material in the typical ISM is rather low, e.g. a few percent. For the same temperature, ultrasmall grains appear to be more efficiently aligned than small grains, with the peak alignment around $10$\AA~due to the resonance relaxation. When accounting for the fact that the temperature of ultrasmall grains is higher with strong fluctuations, the degree of alignment of ultrasmall grains is reduced.

\item[2.] We derived the alignment functions that reproduce the observed polarization curves of the different peak wavelengths $\lambda_{\max}$. We identified that the optical and IR polarization characterized by $\lambda_{\max}$ is mostly produced by RAT-aligned grains with sizes larger than $\sim 0.05\mum$, while the UV polarization is produced by both the $a>0.05\mum$ grains and the $a<0.05\mum$ grains. The sightlines with lower $\lambda_{\max}$ require the higher degrees of alignment of small grains to reproduce the observational data.

\item[3.] We showed that the excess UV continuum polarization relative to the Serkowski law for the sightlines with low $\lambda_{\max}$ ($\lambda_{\max}< 0.55\mum$) can be reproduced by the enhanced paramagnetic alignment of small silicate grains, which higher efficiency arises from the increased magnetic field strength.

\item[4.] We suggested a novel method to measure the strength of magnetic fields based on UV and optical polarization observations. Applying our technique for three sightlines with maximum polarization efficiency, we estimated the upper limit of magnetic field $B\sim 10\mu$G for the typical diffuse ISM of $\lambda_{\max}=0.55\mum$ and larger magnetic fields for the sightlines with $\lambda_{\max}\le 0.53\mum$, assuming oblate spheroid with axial ratio $r=2$ for interstellar grains and average ISRF. Higher magnetic fields are estimated if the oblate spheroid with axial ratio $r=1.5$ is assumed. This technique is complementary to that by Chandrasekhar and Fermi for obtaining a reliable measure of interstellar magnetic fields using dust polarimetry.

\item [5.] We found that the degree of alignment of PAHs required to reproduce the $2175$\AA~polarization feature in HD197770 as derived in \cite{Hoang:2013dw} can be fulfilled by resonance paramagnetic relaxation with the interstellar magnetic field $B\sim 5\mu$G.

\end{itemize}

\acknowledgments
We thank the anonymous referee for valuable comments and suggestions that improved our paper. T.H. is supported by Alexander von Humboldt Fellowship at the Ruhr-Universit$\ddot{\rm a}$t Bochum. P.G.M. acknowledges the support from the Natural Sciences and Engineering Research Council of Canada (NSERC). A.L. acknowledges the financial support of NASA grant NNX11AD32G and the Center for Magnetic Self-Organization.

\appendix
\section{A. Collisional damping times}\label{apdx:A}

The process of gas-grain collisions consists of the sticking collisions followed by the evaporation of molecules from the grain surface. In the grain frame of reference, the mean torque arising from the sticking collisions on an axisymmetric grain rotating around its symmetry axis $\ahat_{1}$ tends to zero when averaged over grain revolving surface. On the other hand, the evaporation induces a non-zero mean torque, which is parallel to the rotation axis (see \citealt{1993ApJ...418..287R}). The damping times for the rotation parallel and perpendicular to the grain symmetry axis $\ahat_{1}$ were derived in \cite{Lazarian:1997p5348}. Basically, the collisional damping time for the rotation along an axis is given by
\bea
\frac{\langle \Delta J_{i}^{b}\rangle}{\Delta t}=-\frac{J_{i}^{b}}{\tau_{\H,i}} {~\mbox {for}~ }i=x,y,z,\label{eq:dJgasdt}
\ena
where the superscript $b$ indicates the grain body system $\ahat_{1}\ahat_{2}\ahat_{3}$, $x,y,z$ denote the components of $J_{i}$ along $\ahat_{2}\ahat_{3}\ahat_{1}$, $\tau_{\H,x}=\tau_{\H,y}\equiv \tau_{\H,\perp}$, and $\tau_{\H,z}=\tau_{\H,\|}$. $\tau_{\H,\|}$ and $\tau_{\H,\perp}$ are given by Equations (\ref{eq:tauHx}) and (\ref{eq:tauHy}). 

Usually, we represent grain angular momentum $J$ in units of the thermal angular momentum and the gaseous damping time. For oblate spheroid, the thermal angular momentum is given by
\bea
J_{\th}=\sqrt{I_{\|}k_{\B}T_{\gas}}=\sqrt{\frac{8\pi\rho a^{5}s}{15}k_{\B}
T_{\gas}}\approx 5.89\times 10^{-20} a_{-5}^{5/2}\hat{s}^{1/2}\hat{\rho}^{1/2}
\hat{T}_{\gas}^{1/2} {\g\cm}^{2}{\rad \s}^{-1},\label{eq:Jth}
\ena
where $\hat{s}=s/0.5$ with $s=a_{1}/a_{2}$ and $\hat{T}_{\gas}=T_{\gas}/100\K$. 

The thermal angular velocity is equal to
\bea
\omega_{\th}=\left(\frac{2k_{\B}T_{\gas}}{I_{\|}}\right)^{1/2}&\approx&1.85\times 10^{5}\hat{s}^{1/2}a_{-5}^{-5/2}\hat{T}_{\gas}^{1/2}\hat{\rho}^{-1/2} \s^{-1}.
\label{eq:ome_th}
\ena

The geometrical factors in Equations (\ref{eq:tauHx}) and (\ref{eq:tauHy}) are given by
\bea
\Gamma_{\|}=\frac{3}{16}\left[3+4(1-e^2)g(e)-e^{-2}(1-(1-e^2)^2)g(e)\right],\label{eq:Gam_par}\\
\Gamma_{\perp}=\frac{3}{32}\left[7-e^2+(1-e^2)^{2}g(e)+(1-2e^2)(1+e^{-2}
[1-(1-e^2)^2)g(e)])\right],\label{eq:Gam_per}
\ena
where $e=\sqrt{1-s^{2}}$ and 
\bea
g_{e}=\frac{1}{2e}\ln\left(\frac{1+e}{1-e}\right).\label{eq:ge}
\ena

\section{B. Diffusion coefficients for magnetic alignment}\label{sec:Bm}
\cite{1951ApJ...114..206D} derived the mean torque for rotational damping by paramagnetic relaxation. In dimensionless units of 
$\tau_{\gas}$, the drifting components in the inertial coordinate system are given by
\bea
A_{\mag,x}=-Z(\theta)\delta_{\mag}J_{x},~A_{\mag,x}=-Z(\theta)\delta_{\mag}J_{y},~A_{\mag,z}=0,
\ena
where $\delta_{\mag}=\tau_{\gas}/\tau_{\mag}$ with $\tau_{\mag}$ being the magnetic alignment timescale due to paramagnetic and resonance paramagnetic relaxation given by Equation (\ref{eq:tau_mag}), and
\bea
Z(\theta)=1+(h-1)\sin^{2}\theta,
\ena
is a correction term for the spheroidal grain shape from its sphere.
 
In addition to the rotational damping, the paramagnetic relaxation also induces rotational excitation, which is a direct result from the principle of detailed balance, i.e., the probability current at each point in phase space tends to vanish
in thermal dynamic equilibrium (see \citealt{Jones:1967p2924}; RL99). Thus, one can obtain the excitation coefficient as follows:
\bea
A_{\mag,x}f(J)-\frac{1}{2}\frac{\partial}{\partial J_{x}}\left(B_{\mag,xx}f(J)\right)=0,
\ena
 where $f=C\exp\left(\frac{ZJ^{2}}{2T_{\d}/T_{\gas}}\right)$ (see also \citealt{Jones:1967p2924}).

Following RL99, one obtain, \footnote{There is a typo in Eq. (3-21) of RL99 for which the correct form should not have the last term of $\left(T_{\d}/T_{\gas}\right)\delta_{\mag}$. Our expressions differ from those of RL99 by a factor 2 because we adopted the normalized units $J_{\th}=\left(I_{\|}k_{\B}T_{\gas}\right)^{1/2}$.}
\bea
B_{\mag,xx}=\frac{T_{\d}}{T_{\gas}}\delta_{\mag},~ B_{\mag,yy}=B_{\mag,xx},~B_{\mag,zz}=0.
\ena

\section{C. Transformation of coordinate systems}\label{sec:Bdiff}
\begin{figure*}
\centering
\includegraphics[width=0.3\textwidth]{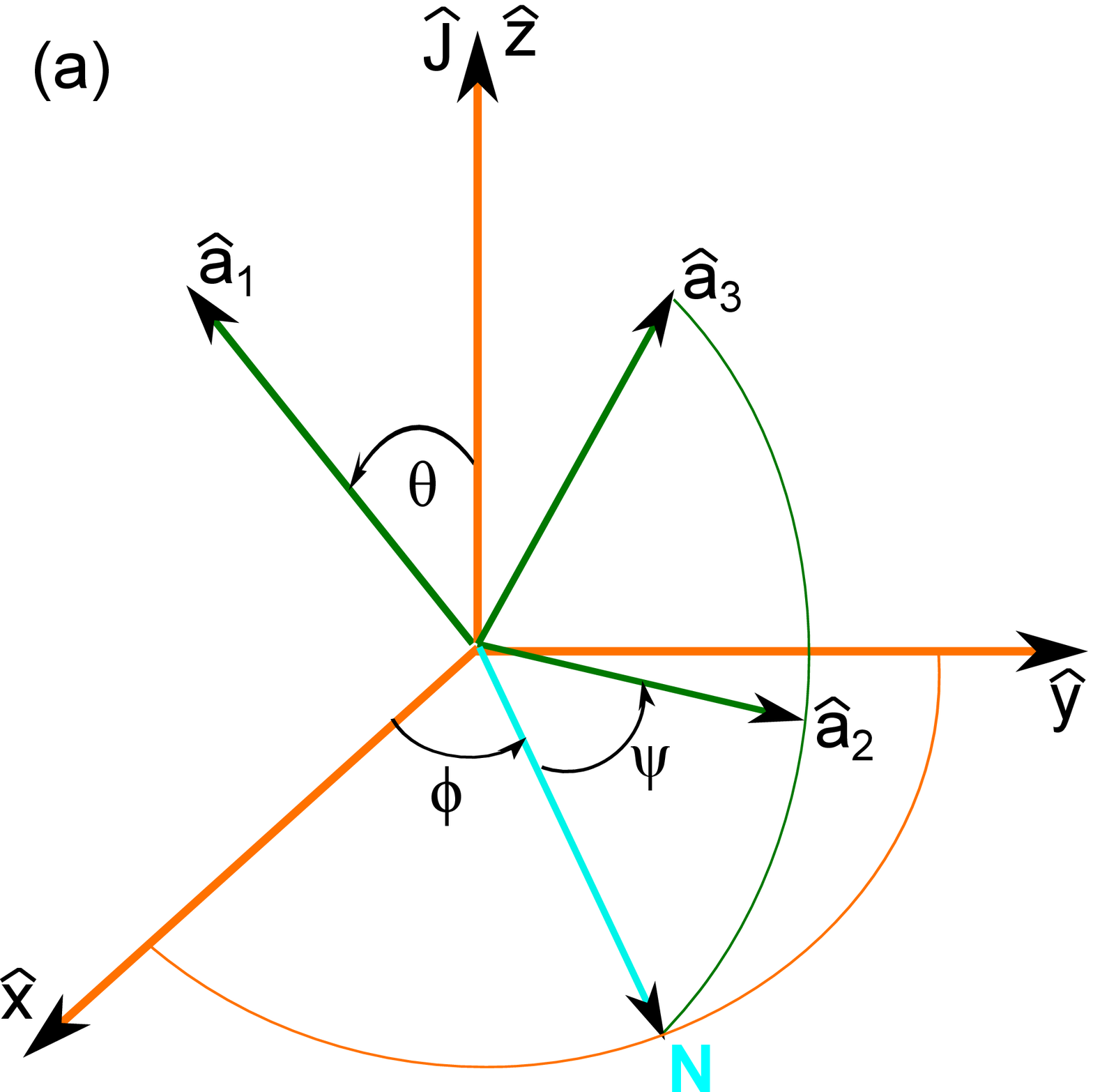}
\includegraphics[width=0.35\textwidth]{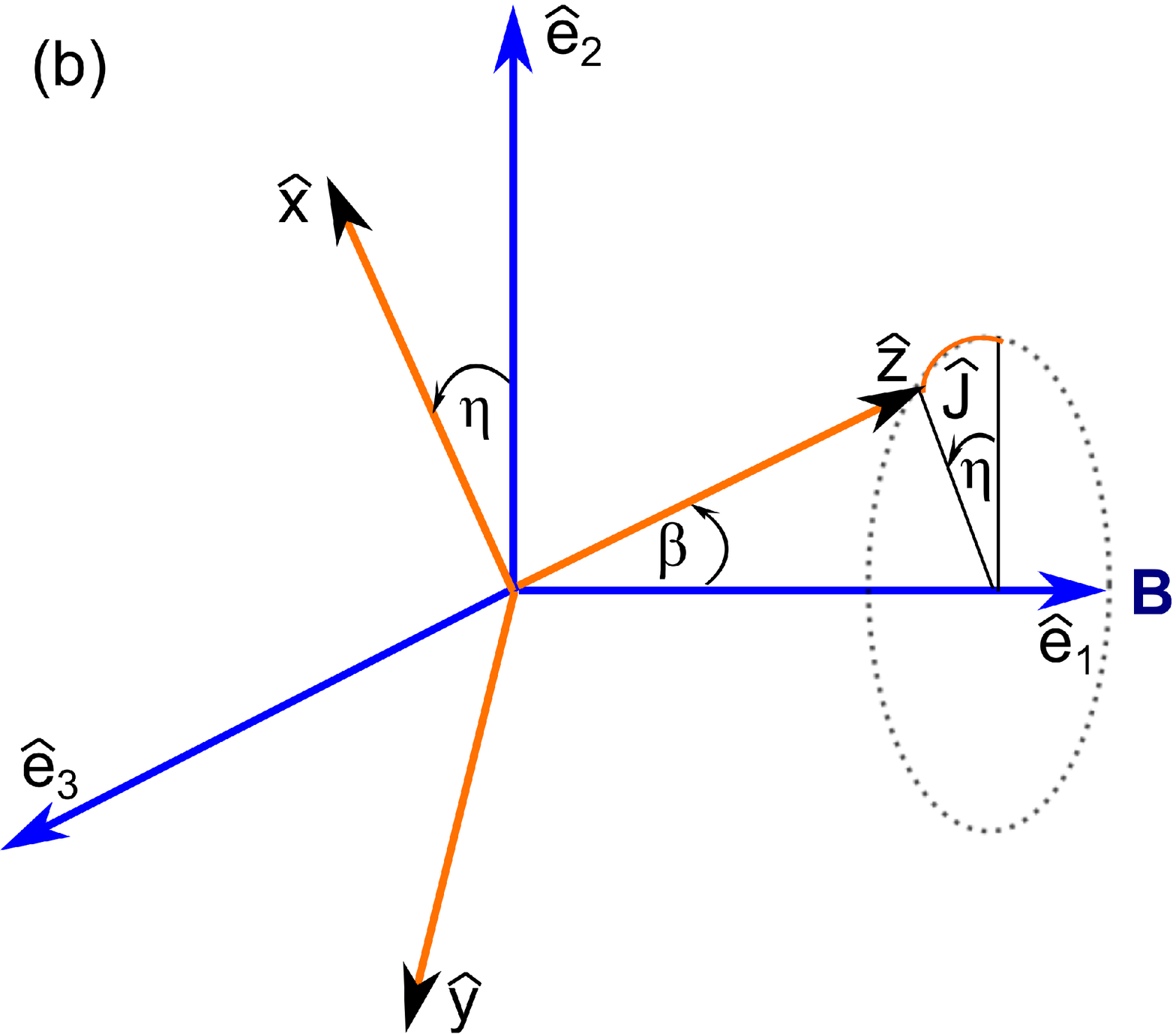}
\caption{Coordinate systems used for calculations. (a): Orientation of grain principal axes in the coordinate system $\xhat\yhat\zhat$ with $\zhat$ parallel to the grain angular momentum $\bJ$. (b): orientation of $\bJ$ in the inertial coordinate system $\ehat_{1}\ehat_{2}\ehat_{3}$ with $\ehat_{1}$ parallel to the magnetic field $\Bv$.}
\label{fig:frames}
\end{figure*}

Damping coefficient $ A_{i}=\langle {\Delta J_{i}}/{\Delta t}\rangle$ and diffusion coefficients  
$B_{ij}=\langle {\Delta J_{i}\Delta J_{j}}/{\Delta t}\rangle$ are usually derived in the body coordinate system, while we are interested in the evolution of grain angular momentum in the inertial coordinate system. Let us define an inertial coordinate system $\ehat_{1}\ehat_{2}\ehat_{3}$ in which  the direction $\bJ$ is described by the angle $\beta$ between $\bJ$ with $\ehat_{1}\| \Bv$, and the azimuthal angle $\eta$ (see Figure \ref{fig:frames}(b)). To obtain these coefficients in the lab coordinate system, we first transform the body system $\ahat_{i}$ to the external system $\xhat \yhat \zhat$ (see \ref{fig:frames}(a)). Then, we perform the transformation from $\xhat\yhat\zhat$ system to the inertial system $\ehat_{1}\ehat_{2}\ehat_{3}$. 

In the body system, the damping coefficients are given by
\bea
A_{i}^{b}=\langle \frac{\Delta J_{i}^{b}}{\Delta t}\rangle=
-\frac{J_{i}}{\tau_{{\rm gas},i}}-\frac{J_{i}^{3}}{\tau_{{\rm ed},i}}
\left(\frac{1}{3I_{i}k_{\rm B}T_{\rm gas}}\right),
\ena
where $\tau_{{\gas},i}=F_{{\rm tot},i}/\tau_{\rm H,\|}$ and $i=x,y,z$ with $z\| \ahat_{1}$.

The diffusion coefficients in the grain body system, $B_{ij}^{b}=\langle {\Delta J_{i}^{b}\Delta J_{j}^{b}}/{\Delta t}\rangle$ 
with $B_{ij}^{b}=0$ for $i\ne j$ are related to the excitation coefficients as follows:
\bea
B_{zz}^{b}=B_{\|}=\frac{2I_{\|}\kB T_{\rm gas}}{\tau_{\rm H,\|}}G_{\rm tot,\|},{\rm ~and~~}
B_{xx}^{b}=B_{yy}^{b}=B_{\perp}=\frac{2I_{\perp}\kB T_{\rm gas}}{\tau_{\rm H,\perp}}G_{\rm tot,\perp}.
\ena

The diffusion coefficients in the inertial system $\ehat_{i}$ have components $ B_{11}, B_{22}$ and $B_{33}$, which are denoted by $B_{zz}, B_{xx}, B_{yy}$ for consistency.  Using the method in \cite{Lazarian:1997p5348} to perform the transformations from the body system to inertial system, after averaging over the fast precession of the grain symmetry axis around angular momentum, we obtain 
\bea
B_{zz}&=&B_{\|}\left(\frac{1}{2}
\sin^{2}\theta\sin^{2}\beta+\cos^{2}\theta\cos^{2}\beta\right)+B_{\perp}\left(\frac{1}{2}[
1+\cos^{2}\theta]\sin^{2}\beta+\sin^{2}\theta \cos^{2}\beta\right),\label{eq:Bzz}\\
B_{xx}&=&B_{\|}\left(\frac{1}{2}\sin^{2}\theta[\cos^{2}\eta+\sin^{2}\eta\cos^{2}\beta]
+\cos^{2}\theta \sin^{2}\eta\sin^{2}\beta\right)+B_{\perp}\left(\frac{1}{2}[1+\cos^{2}\theta]
[\cos^{2}\eta+\sin^{2}\eta\cos^{2}\beta]
+\sin^{2}\theta\sin^{2}\eta\sin^{2}\beta\right),\\
B_{yy}&=&B_{\|}\left(\frac{1}{2}\sin^{2}\theta[\sin^{2}\eta+\cos^{2}\eta\cos^{2}\beta]
+\cos^{2}\theta \sin^{2}\eta\sin^{2}\beta\right)+B_{\perp}\left(\frac{1}{2}[1+\cos^{2}\theta]
[\sin^{2}\eta+\cos^{2}\eta\cos^{2}\beta]
+\sin^{2}\theta\sin^{2}\eta\sin^{2}\beta\right),\label{eq:Byy}
\ena
where $\beta$ is the angle between $\bJ$ and $\ehat_{1}$, and $\eta$ is the azimuthal angle of $\bJ$ in the inertial system $\ehat_{i}$.

In the presence of fast internal fluctuations, we need to average the damping and diffusion coefficients over $\theta$. Therefore, the terms containing $\theta$ in above equations are replaced by the averaged values, i.e., $\langle\cos^{2}\theta\rangle=\int_{0}^{\pi} \cos^{2}\theta f_{\rm LTE}(J,\theta) \sin\theta d\theta$,
$\langle\sin^{2}\theta\rangle=\int_{0}^{\pi} \sin^{2}\theta f_{\rm LTE}(J,\theta) \sin\theta d\theta$.

In the presence of ambient magnetic field, the grain angular momentum precesses around $\Bv$ on a timescale $\tau_{\Lar}$ (Equation \ref{eq:tauB}), which is short compared to the dynamical timescales due to gas bombardment, electric dipole emission, and IR emission. Therefore, one can average the damping and diffusion coefficients over the uniform distribution of the precession angle $\eta$. Thus, $\sin^{2}\eta$ and $\cos^{2}\eta$ are replaced by their averaged values equal to $1/2$. In this case, our diffusion coefficients (Eqs \ref{eq:Bzz}-\ref{eq:Byy}) become similar to those in \citep{Lazarian:1997p5348}.

\section{D. Extinction and Polarization}\label{sec:extpol}
\subsection{D1. Dust Extinction and Polarization}

To find the extinction and polarization of background starlight by interstellar grains, let us define an observer's coordinate system in which the sightline is directed along the $Z-$axis, and the $X-$ and $Y-$ axes constitute the sky plane. The polarization of starlight arising from the dichroic extinction by aligned grains in a cell of $dZ$ is computed as
\bea
dp({\lambda})=\frac{d\tau_{X}-d\tau_{Y}}{2}=\int_{a_{\min}}^{a_{\max}}\frac{1}{2}\left(C_{X}-C_{Y}\right)
(dn/da)dadZ,
\label{eq:dplam}
\ena
where $dn/da$ is the grain size distribution function with the lower and upper cutoff $a_{\min}$ and $a_{\max}$, $C_{X}$ and $C_{Y}$ are the grain cross-section along the $X-$ and $Y-$ axes, respectively.

For the case of perfect internal alignment (i.e., grain symmetry axis $\ahat_{1}$ perfectly aligned with its angular momentum), by transforming the grain's reference system to the observer's reference system and taking corresponding weights, we obtain
\bea
C_{X}&=&C_{\perp}-\frac{C_{\pol}}{2}\sin^{2}\beta,\\
C_{Y}&=&C_{\perp}-\frac{C_{\pol}}{2}(2\cos^{2}\beta\cos^{2}\xi+\sin^{2}\beta\sin^{2}\xi),
\ena
where $\xi$ is the angle between the magnetic field assumed to be in the $YZ$ plane and the sky plane, $\beta$ is the angle between the grain angular momentum and the magnetic field, and $C_{\pol}=C_{\|}-C_{\perp}$ is the polarization cross-section for oblate spheroidal grains. By convention, $C_{\|}$ and $C_{\perp}$ are the extinction cross-section for the electric field of incident radiation parallel and perpendicular to the grain symmetry axis, respectively. 

The polarization efficiency then becomes
\bea
C_{X}-C_{Y}=C_{\pol}\frac{\left(3\cos^{2}\beta-1\right)}{2}\cos^{2}\xi.\label{eq:Cpol}
\ena

Taking the average of $C_{X}-C_{Y}$ over the distribution of the alignment angle $\beta$, it yields
\bea
C_{X}-C_{Y}=C_{\pol}\langle Q_{J}\rangle \cos^{2}\xi,\label{eq:Cx-Cy}
\ena
where $Q_{J}=\langle G_{J}\rangle$ is the ensemble average of $G_{J}=\left(3\cos^{2}\beta-1\right)/2$ that describes the alignment of grain angular momentum with the ambient magnetic field.

When the internal alignment is not perfect, following the similar procedure, we obtain
\bea
C_{X}-C_{Y}=C_{\pol}\langle Q_{J}Q_{X} \rangle \cos^{2}\xi \equiv
C_{\pol}R\cos^{2}\xi,\label{eq:Cpol}
\ena
where $R=\langle Q_{J}Q_{X}\rangle$ is the Rayleigh reduction factor (see also RL99).

Let $f=R\cos^{2}\xi$ be the effective degree of grain alignment. Thus, for the case of perpendicular magnetic field, i.e., $\Bv$ lies on the sky plane $f=R$, Equation (\ref{eq:Cpol}) simply becomes $C_{X}-C_{Y}=C_{\pol}f$.

Plugging in Equation (\ref{eq:Cpol}) into this above equation, we obtain
\bea
p({\lambda})=\int dZ\sum_{j=\carb,\sil}\int_{a_{\min}}^{a_{\max}} \frac{1}{2}C_{\pol}^{j}f^{j}(a)
(dn^{j}/da)da,\label{eq:Plam}
\ena
where $f^{j}(a)$ denotes the alignment function of grain specie $j$ of size $a$.

The extinction in units of magnitude is defined by
\bea
A({\lambda})&=&2.5{\rm log}_{10}\left(\frac{F_{\lambda}^{\obs}}{F_{\lambda}^{\star}}\right),\nonumber\\
&=&1.086\tau_{\lambda}=1.086\int dZ \sum_{j=\carb,\sil}\int_{a_{\min}}^{a_{\max}} C_{\ext}^{j}
(dn^{j}/da)da,\label{eq:Aext}
\ena
where $F_{\lambda}^{\star}$ is the intrinsic flux from the star, $F_{\lambda}^{\obs}=F_{\lambda}^{\star}e^{-\tau{_\lambda}}$ is the observed flux, and $\tau_{\lambda}$ is the optical depth.

Frequently, it is more convenient to represent the polarization (extinction) through the polarization (extinction) cross-section. Hence, the above equations can be rewritten as
\bea
p({\lambda})=\sigma_{\pol}(\lambda)\times N_{\H},\\
A({\lambda})=\sigma_{\ext}(\lambda)\times N_{\H},
\ena
where $N_{\H}(\cm^{-2})$ is the column density and $\sigma_{\ext}$ and $\sigma_{\pol}$ in units of $\cm^{2}~\H^{-1}$ are the dust extinction cross-section and dust polarization cross-section, respectively.

We take $C_{\ext}(a,\lambda)$ and $C_{\pol}(a,\lambda)$ computed for silicate and carbonaceous grains in \cite{Hoang:2013dw}.

\section{E. Nonlinear Least chi-square fitting}\label{sec:chisq}

Following \cite{1995ApJ...444..293K}, we find the grain size distribution and alignment function by minimizing an objective function $\chi^{2}$, which is constructed as follows:
\bea
\chi^{2}=\chi_{\ext}^{2}+\chi_{\pol}^{2}+\chi_{\rm con}^{2},\label{eq:chisq}
\ena
where
\bea
\chi_{\ext}^{2}=w_{\ext}\sum_{i=0}^{N_{\lambda}-1}\left[A_{\mod}(\lambda_{i})-A_{\obs}(\lambda_{i})\right]^{2},\\
\chi_{\pol}^{2}=w_{\pol}\sum_{i=0}^{N_{\lambda}-1}\left[p_{\mod}(\lambda_{i})-p_{\obs}(\lambda_{i})\right]^{2},
\ena
with $w_{\ext}$ and $w_{\pol}$ being the fitting weights for the extinction and polarization, respectively. Here, the summation is performed over $N_{\lambda}$ wavelength bins. For this study, we adopt $N_{a}=100$ size bins from $a=3.56$\AA~to $1\mum$ ~and $N_{\lambda}=100$ from $\lambda=0.125\mum$ to $2.5\mum$. The last term $\chi_{\rm con}^{2}=\sum \Psi^{2}$ contains the constraints of the fitting model, which are similar to Equations (A5)-(A9) in \cite{2006ApJ...652.1318D}. Below we provide them here for consistency.

\bea
\Psi_{2N_{\lambda}+j+2}&=&\frac{\alpha_{5}}{\left(N_{a}-1\right)^{1/2}}
\left(\min \left[\left(\frac{d\ln f}{du}\right)_{j+1/2},0\right]\right)^{2},\label{eq:psi5}\\
\Psi_{2N_{\lambda}+N_{a}+2}&=&\alpha_{6}{\max\left[f(a_{N_{a}})-1,0\right]}^{2},\label{eq:psi6}\\
\Psi_{2N_{\lambda}+N_{a}+1+j}&=&\frac{\alpha_{7}}{\left(N_{a}-1\right)^{1/2}}\left(\frac{d^{2}y_{\sil}}{du^{2}}\right),a=a_{j}, j=2,...,N_{a}-1,\label{eq:psi7}\\
\Psi_{2N_{\lambda}+2N_{a}-1+j}&=&\frac{\alpha_{8}}{\left(N_{a}-1\right)^{1/2}}\left(\frac{d^{2}y_{\carb}}{du^{2}}\right), a=a_{j}, j=2,...,N_{a}-1,\label{eq:psi8}\\
\Psi_{2N_{\lambda}+3N_{a}-3+j}&=&\frac{\alpha_{9}}{\left(N_{a}-1\right)^{1/2}}\left(\frac{d^{2}\ln f}{du^{2}}\right), a=a_{j}, j=2,...,N_{a}-1,\label{eq:psi9}\\
\ena
where $du=\ln a_{j+1}-\ln a_{j}$, $\left(df/du\right)_{j+1/2}=\left(f_{j+1/2}-f_{j}\right)/\Delta u$ and $\left(d^{2}f/du^{2}\right)_{j}=\left(f_{j+1}+f_{j-1}-2f_{j}\right)/\left(\Delta u\right)^2$, and $\alpha_{5}-\alpha_{9}$ are weights, which are quite arbitrary.

The objective functions for extinction and polarization are different from those of \cite{2006ApJ...652.1318D} in a sense that our objective functions are constructed from the difference between the model and observation.

We find the minimum $\chi^{2}$ using the Monte-Carlo direct search method in which the fitting process is iterated until the convergence criterion is achieved.

\bibliography{ms.bbl}

\end{document}